\documentclass{llncs}
\usepackage{amsmath,amssymb,amsfonts,enumerate,bbm}
\usepackage{ifthen}
\usepackage{graphicx}

\usepackage{subfigure}

\usepackage{color,soul}

\long\def\commabs #1\commabsend{}
\long\def\commful #1\commfulend{#1}
\long\def\comment #1\commentend{}

\def\dnsparagraph#1{\par\vspace{2pt}\noindent{\bf #1}.}

\newtheorem{observation}[theorem]{Observation}
\newtheorem{fact}[theorem]{Fact}
\def\deg{\mbox{\tt deg}}

\def\nearest{\mbox{\tt nearest}}

\def\Cost{\mbox{\tt Cost}}
\def\Value{\mbox{\tt Val}}
\newcommand{\dist}{\mbox{\rm dist}}

\def\inline#1:{\par\vskip 7pt\noindent{\bf #1:}\hskip 10pt}
\def\Proof{\par\noindent{\bf Proof:~}}
\def\blackslug{\hbox{\hskip 1pt \vrule width 4pt height 8pt
    depth 1.5pt \hskip 1pt}}
\def\QED{\quad\blackslug\lower 8.5pt\null\par}
\def\CLUSTER{\mbox{\sf Cluster}}
\newcommand{\HYBRID}[0]{\mbox{\sf ConsHybrid}}
\newcommand{\SOURCEWISES}[0]{\mbox{\sf ConsSWSpanner}}
\def\splenone{\ell_t}
\def\splentwo{\ell_t}
\begin{document}

%%%%%%%%%%%%%%%%%%%%%%
\title{Bypassing Erd\H{o}s' Girth Conjecture: Hybrid Stretch and Sourcewise Spanners}

%\author{
%Merav Parter
%\thanks{Department of Computer Science and Applied Mathematics,
%The Weizmann Institute of Science, Rehovot, Israel.
%E-mail: {\tt \{merav.parter\}@ weizmann.ac.il}.
%Supported in part by the Israel Science Foundation (grant 894/09),
%the United States-Israel Binational Science Foundation (grant 2008348),
%the Israel Ministry of Science and Technology (infrastructures grant),
%and the Citi Foundation.
%Recipient of the Google European Fellowship in distributed computing;
%research is supported in part by this Fellowship.}
%}
%\date{}

\author{
Merav Parter}
\institute{
The Weizmann Institute of Science, Rehovot, Israel.
\email{merav.parter@weizmann.ac.il}
\thanks{Recipient of the Google European Fellowship in distributed computing; research supported in part by this Fellowship. Supported in part by the Israel Science Foundation (grant 894/09), United States-Israel Binational Science Foundation (grant 2008348), Israel Ministry of Science and Technology (infrastructures grant), and Citi Foundation.}}
\maketitle
%%%%%%%%%%%%%%%%%%%%%%
%\pagenumbering{gobble}

\begin{abstract}
An $(\alpha,\beta)$-spanner of an $n$-vertex graph $G=(V,E)$ is a subgraph $H$ of $G$ satisfying that $\dist(u, v, H) \leq \alpha \cdot \dist(u, v, G)+\beta$ for every pair $(u, v)\in V \times V$, where $\dist(u,v,G')$ denotes the distance between $u$ and $v$ in $G' \subseteq G$. It is known that for every integer $k \geq 1$, every graph $G$  has a polynomially constructible $(2k-1,0)$-spanner of size $O(n^{1+1/k})$. This size-stretch bound is essentially optimal by the girth conjecture. Yet, it is important to note that any argument based on the girth only applies to \emph{adjacent vertices}. It is therefore intriguing to ask if one can ``bypass" the conjecture by settling for a multiplicative stretch of $2k-1$ only for \emph{neighboring} vertex pairs, while maintaining a strictly \emph{better} multiplicative stretch for the rest of the pairs. We answer this question in the affirmative and introduce the notion of \emph{$k$-hybrid spanners}, in which non neighboring vertex pairs enjoy a \emph{multiplicative} $k$-stretch and the neighboring vertex pairs enjoy a \emph{multiplicative} $(2k-1)$ stretch (hence, tight by the conjecture). We show that for every unweighted $n$-vertex graph $G$ with $m$ edges, there is a (polynomially constructible) $k$-hybrid spanner with $O(k^2 \cdot n^{1+1/k})$ edges. This should be compared against the current best $(\alpha,\beta)$ spanner construction of \cite{BSADD10} that obtains $(k,k-1)$ stretch with $O(k \cdot n^{1+1/k})$ edges.
\indent An alternative natural approach to bypass the girth conjecture
is to allow ourself to take care only of a subset of pairs $S \times V$ for a given subset of vertices $S \subseteq V$  referred to here as \emph{sources}. Spanners in which the distances in $S \times V$ are bounded are referred to as \emph{sourcewise spanners}. Several constructions for this variant are provided (e.g., multiplicative sourcewise spanners, additive sourcewise spanners and more).
\end{abstract}
%\newpage
%\newpage
%\pagenumbering{arabic}
%(i) for a given subset $S \subseteq V$ of size $O(n^{\varepsilon})$, we construct $2k$ sourcewise hybrid spanners with $O(n^{1+\varepsilon/(k+1)})$ edges, (ii) we provide a tight construction of sourcewise $(1,2)$ emulators with $O(n^{1+\varepsilon/2})$ edges and (iii) we study
%
%turn the $V \times V$
%
%For a given subset of vertices $S \subseteq V$, referred to as \emph{sources}, the spanner aims to bound only the $S \times V$ distances. Such subgraphs are called \emph{sourcewise spanners}.
%We provide spanners constructions that are parameterized by the number of sources. Specifically, we obtain a %sourcewise $2k$-hybrid spanner with $O(n^{1+\varepsilon/(k+1)})$ edges and an $2k$-additive %sourcewise spanner with $O(n^{1+(k\varepsilon+1/(2k+1)})$ edges. We complement this result by establishing a lower bound for sourcewise $(1,2k)$ spanners. In particular, we parameterize the lower bound construction of Woodruff for $(1,2k)$ spanners by the number of sources $|S|=n^{\varepsilon}$.
%Finally, we study the sourcewise variants of approximated distance oracles and additive emulators\footnote{For a graph $G=(V,E)$, an $(\alpha, \beta)$ emulator $H=(V,F)$ satisfies $\dist(u, v, G) \leq \dist(u, v, H) \leq \alpha \cdot \dist(u, v, G)+\beta$ but need not be subgraph of $G$ and can be weighted}. We achieve a tight construction of %$(1,2k)$ $S \times V$ emulator with %$O(n^{1+\varepsilon/2})$ edges.
%
\section{Introduction}
\subsection{Motivation}
Graph spanners are sparse subgraphs that faithfully preserve the pairwise distances of a given graph and provide the underlying graph structure in communication networks, robotics, distributed systems and more \cite{Peleg00:book}. The notion of graph spanners was introduced in \cite{PelegSpanner,PelegUllman} and have been studied extensively since. Spanners have a wide range of applications from distance oracles \cite{TZ05,DOBaswana06}, labeling schemes \cite{Preservers} and routing \cite{CyrilPeleg03} to solving linear systems \cite{EmekLowStretch} and spectral sparsification \cite{KapralovITCSSpanner}.

Given an undirected unweighted $n$-vertex graph $G=(V,E)$, a subgraph $H$ of $G$ is said to be a $k$-spanner if for every pair of vertices $(u, v) \in V \times V$ it holds that $\dist(u, v, H) \leq k \cdot \dist(u, v, G)$. It is well known that one can efficiently construct a $(2k-1)$-spanner with $O(n^{1+1/k})$ edges, even for weighted graphs \cite{MultSpanner93,MultWeighBaswana}. This size-stretch ratio is conjectured to be tight based on the girth\footnote{The girth is the smallest cycle length.} conjecture of Erd\H{o}s \cite{Erdos}, which says that there exist graphs with $\Omega(n^{1+1/k})$ edges and girth  $2k+1$. If one removes an edge in such a graph, the distance between the edge endpoints increases from $1$ to $2k$, implying that any $\alpha$-spanner for $\alpha \leq 2k-1$ has $\Omega(n^{1+1/k})$ edges. This conjecture has been resolved for the special cases of $k=1,2,3,5$ \cite{Wenger91}.
\par Although the girth conjecture exactly characterizes the optimal tradeoff between sparseness and multiplicative stretch, it applies only to adjacent vertices (i.e., removing an edge $(u,v)$ from a large cycle causes distortion to the edge endpoints).
Indeed, Elkin and Peleg \cite{PelegElkinMixedSpanner} showed that the girth bound (on multiplicative distortion) fails to hold even for vertices at distance $2$.
This limitation of the girth argument motivated distinguishing between \emph{nearby} vertex pairs and ``sufficiently distant" vertex pairs. This gave raise to the development of $(\alpha, \beta)$-spanners which distort distances in $G$ up to a multiplicative factor of $\alpha$ and an additive term $\beta$ \cite{PelegElkinMixedSpanner}. Formally, for an unweighted undirected graph $G=(V,E)$, a subgraph $H$ of $G$ is an $(\alpha, \beta)$-spanner iff $\dist(u, v, H) \leq \alpha \cdot \dist(u, v, G)+ \beta$ for every $u, v \in V$. Note, that an $(\alpha, \beta)$-spanner makes an implicit distinction between nearby vertex pairs and sufficiently distant vertex pairs. In particular, for ``sufficiently distant" vertex pairs the $(\alpha, \beta)$-spanner behaves similar to a pure multiplicative spanner, whereas  for the remaining vertex pairs, the spanner behaves similar to an additive spanner \cite{AddSpanner93}.
The setting of $(\alpha, \beta)$-spanners has been widely studied for various distortion-sparseness tradeoffs \cite{ElkinMixSp05,ThorupSublinear06,PelegElkinMixedSpanner,BSADD10}.
For example, \cite{PelegElkinMixedSpanner} gave a construction for $(k -1, 2k -O(1))$-spanners with size $O(k\cdot n^{1+1/k})$, with a number of refinements for short distances, and showed that for any $k \geq 2$ and $\epsilon > 0$, there exist $(1 + \epsilon, \beta)$-spanners with size $O(\beta \cdot n^{1+1/k})$, where $\beta$ depends on $\epsilon$ and $k$ but independent on $n$, implying that the size can be driven close to linear in $n$ and the multiplicative stretch close to $1$, at the cost of a large additive term in the stretch. Thorup and Zwick designed  $(1 + \epsilon, \beta)$-spanners with $O(k \cdot n^{1+1/k})$ edges, with a multiplicative distortion that tends to $1$ as the distance increases \cite{ThorupSublinear06}.
\par The best $(\alpha, \beta)$ spanner construction is due to \cite{BSADD10} which achieves stretch of $(k, k-1)$ with $O(k \cdot n^{1+1/k})$ edges, hence providing multiplicative stretch $2k-1$ for neighboring vertices (which is the best possible by Erd\H{o}s' conjecture) and a multiplicative stretch at most
$3k/2$ for the remaining pairs.

Although $(\alpha, \beta)$-spanners make an (implicit) distinction between ``close" and ``distant" vertex pairs, as the girth argument holds only for vertices at distance $1$, it seems that a tighter bound on the behavior of spanners may be obtained. In particular, it seems plausible that the multiplicative factor of $k$ using $O(n^{1+1/k})$ edges, is not entirely unavoidable for non-neighboring vertex pairs, while providing multiplicative stretch of $2k-1$ for the neighboring vertex pairs.
%We show that one can bypass the conjecture without introducing an additional parameter (e.g., $\beta$). Specifically, to ``bypass" the conjecture it is sufficient to have a \emph{single} stretch parameter $k$ that has \emph{two} types of interpretations:
%\textbf{MP: maybe we should not stick to the additive vrs. %multiplicative but rather to multiplicative factor of $2k-1$ vrs. multiplicative stretch $k$?}\\
%an additive stretch of $2(k-1)$ for neighboring vertex pairs and a \emph{multiplicative} stretch of $k$ for the remaining vertex pairs.
The current paper confirms this intuition by introducing the notion of $k$-hybrid spanners, namely, subgraphs $H \subseteq G$ that obtain multiplicative stretch $2k-1$ for neighboring vertices, i.e.,  $\dist(u, v, H) \leq (2k-1) \cdot\dist(u, v, G)$ for every $(u, v) \in E(G)$ and  multiplicative stretch $k$ for the remaining vertex pairs, i.e., $\dist(u, v, H) \leq k \cdot \dist(u, v, G)$ for every $(u, v) \notin E(G)$.
%More precisely, in $(\alpha, \beta)$-spanners, vertices of distance $d$ in $G$ have distance at most $f(d)=\alpha \cdot d+\beta$ and in $k$-hybrid spanners, their distance is at most $f'(d)=\max\{k \cdot d, d+2(k-1)\}$.
Hence, hybrid spanners seem to pinpoint the minimum possible relaxation of the stretch requirement in spanners graphs so that the girth conjecture lower bound can be bypassed.
The presented $k$-hybrid spanner with $O(k^2 \cdot n^{1+1/k})$ edges can be contrasted with several existing spanner constructions, e.g, $k$-spanners with $O(n^{1+2/(k+1)})$ edges (in which multiplicative stretch $k$ is guaranteed also to neighboring pairs), the $\Omega(k\cdot n^{1+1/k})$ lower-bound graph construction for $(2k-1)$-additive spanners, and to the $(k,k-1)$ spanner construction of \cite{BSADD10} with $O(k^2 \cdot n^{1+1/k})$ edges.
%that using the same number of edges provides a path of length $3 \cdot d+2$ for any vertex pair at distance $d$. As an illustration, fixing the number of edges to $O(n^{4/3})$ results in $3$ hybrid spanner with multiplicative stretch $5$ for neighboring vertex pairs and a multiplicative stretch of $3$ for the remaining pairs. This should be compared against two other spanner constructions with the same order of number of edges, namely, the standard $5$-multiplicative spanner and the $(3,2)$ spanner construction of \cite{BSADD10}.
%
%
%
\par An alternative approach to bypass the conjecture is by focusing on a subset of pairs in $V \times V$. Following \cite{PairwisePresElkin06,PettieLowDist,CGK13,PairwiseICALP13}, we relax the requirement that small stretch in the subgraph must be guaranteed for \emph{every} vertex pair from $V \times V$. Instead, we require it to hold only for pairs of vertices from a subset of $V \times V$.
Specifically, given a subset of vertices $S \subseteq V$, referred to here as \emph{sources}, our spanner $H$ aims to bound only the distances between pairs of vertices from $S \times V$. For any other pair outside $S \times V$, the stretch in $H$ can be arbitrary.
%We call such a subgraph $H$ an \emph{$(\alpha, \beta,S)$-sourcewise spanner}.
%
%\textbf{MP: Requires rewriting if the hybrid does not turn into sourcewise...}
%\par Combining the two approaches for bypassing the conjecture yields the notion of \emph{hybrid sourcewise spanners}. We show that for a stretch parameter $2k$, for an integer $k \geq 1$, given a subset of sources $S \subseteq V$ of cardinality $O(n^{\varepsilon})$, there exists a (polynomial time constructible) sourcewise hybrid spanner with $O(k^2 \cdot n^{1+\varepsilon/(k+1)})$ edges.
%
%
%This paper gives the system designer an alternative compromise, where, for example, instead of increasing the stretch by a factor of $2$ he can consider only a subset $S$ of $\sqrt{n}$ vertices (as sources) and maintain the \emph{same} stretch guarantee as before, at the expense of maintaining this guarantee only for the pairs in $S \times V$. In other words, when fixing on some initial stretch value $2k$, our upper bound construction implies that the size-stretch tradeoff is the same as the size-(number of sources) tradeoff: \emph{increasing} the initial stretch of $2k$ by an integral factor of $c\geq 1$ has the same effect on the size of the new spanner as taking a subset $S \subseteq V$ of size $|S|=|V|^{1/c}$, i.e., \emph{decreasing} the number of sources by a factor of $|V|^{1-1/c}$. The first compromise results in an
%$(c \cdot (2k),0)$ $V \times V$ spanner and the second compromise results in a $(2k,0)$ $S \times V$ spanner, both with $O(n^{1+1/(c \cdot k}))$ edges.
%
\par On the lower bound side, Woodruff \cite{Woodruff06} proved, independently of the Erd\H{o}s' conjecture, the existence of graphs for which any spanner of size $\Omega(k^{-1} n^{1+1/k})$ has an additive stretch of at least $2k-1$.
%This construction generalizes to \emph{graph emulators} \cite{EmulatorDor}. An $(\alpha, \beta)$-emulator $H'=(V,F)$ extends the notion of $(\alpha, \beta)$-spanner $H$: it provides the same stretch guarantee for every pair $\langle u, v \rangle \in V \times V$ (e.g., $\dist(u, v, G)\leq \dist(u, v, H') \leq \alpha \cdot \dist(u, v, G)+\beta$) but it might be weighted and need not be a subgraph of $G$.
%Currently, additive spanner constructions are known only for additive stretch of $2,4$ and $6$ \cite{Aing99,4Add,BSADD10}. Tight emulator construction is known only for the case of additive stretch $2$ and $4$ (\cite{Woodruff06,EmulatorDor}).
Although sourcewise additive spanners have been studied by \cite{PettieLowDist,CGK13,PairwiseICALP13}, currently there are no known lower bound constructions for this variant. We generalize Woodruff's construction to the sourcewise setting, providing a graph construction whose size has a smooth dependence with the number of sources.

\subsection{Related Works}
The notion of a sparse subgraph that preserves distances only for a subset of the $V\times V$ pairs has been initiated by Bollob\'{a}s, Coopersmith and Elkin \cite{Preservers}, who studied \emph{pairwise preservers}, where the input is a graph $G=(V,E)$ along with a subset of vertex pairs $\mathcal{P} \subseteq V \times V$ and the problem is to construct a sparse subgraph $H$ such that the $u-v$ distance for each $(u, v) \in \mathcal{P}$ is exactly preserved, i.e., $\dist(u, v, H)=\dist(u, v, G)$ for every $(u, v) \in \mathcal{P}$.  They showed that one can construct a pairwise preserver with $O(\min\{|\mathcal{P}| \cdot \sqrt{n}, n \cdot \sqrt{|\mathcal{P}|}\})$ edges. At the end of their paper, they raised the question of constructing sparser subgraphs where distances between pairs in $\mathcal{P}$ are \emph{approximately} preserved, or in other words, the problem of constructing sparse $\mathcal{P}$-spanners. Pettie \cite{PettieLowDist} studied a certain type of $\mathcal{P}$-spanners, namely, additive sourcewise spanners. In this setting, one is given an unweighted graph $G=(V,E)$ and a subset of vertices $S \subseteq V$, termed as \emph{sources}, whose size is conveniently parameterized to be $|S|=n^{\varepsilon}$, and the goal is to construct a sparse spanner $H$ that maintains an additive approximation for the $S \times V$ distances. He showed a construction of
$O(\log n)$-additive sourcewise spanners of size $O(n^{1+\varepsilon/2})$. Cygan et al. recently showed a stretch-size bound for $2k$-additive sourcewise spanners with $O(n^{1+(\varepsilon k +1)/(2k+1)})$ edges. The specific case of $k=1$ has been studied recently by \cite{PairwiseICALP13}, providing a $2$-additive sourcewise spanner with $\widetilde{O}(n^{5/4+\varepsilon/4})$ edges where $|S|=n^{\varepsilon}$.
%\par To the best of our knowledge, there is no lower bound for the sourcewise variant of additive spanners.
%For the standard $V \times V$ case, Woodruff \cite{Woodruff06} provides a lower bound construction that achieves the same stretch-size bound as given by Erd\H{o}s' conjecture.
\par Upper bounds for spanners with constant stretch
are currently known for but a few stretch values.
A $(1,2)$ spanner with $O(n^{3/2})$ edges is presented in \cite{Aing99},
a $(1,6)$ spanner with $O(n^{4/3})$ edges is presented in \cite{BSADD10}, and
a $(1,4)$ spanner with $O(n^{7/5})$ edges is presented in \cite{4Add}. The latter two constructions use the {\em path-buying} strategy, which is adopted in our additive sourcewise construction. Dor et al. \cite{EmulatorDor} considered additive emulators, which may contain additional (possibly weighted) edges. They showed a construction of $4$-additive emulator with $O(n^{4/3})$ edges.
Finally, a well known application of $\alpha$-spanners is \emph{approximate distance oracles} \cite{TZ05,PatrascuByondT,DOBaswana06,BaswanaK06,Naor93}.
The sourcewise variant, namely, \emph{sourcewise approximate distance oracle} was devised by \cite{Roditty05}. For a given input graph $G=(V,E)$ and a source set $S \subseteq V$, \cite{Roditty05} provides a construction of a distance oracle of size $O(n^{1+\varepsilon/k})$ where $\varepsilon=\log|S|/\log n$
such that given a distance query $(s, v) \in S \times V$ returns in $O(k)$ time a $(2k-1)$ approximation to $\dist(s, v, G)$.

\subsection{Contributions}

In this paper we initiate the study of $k$-hybrid spanners which seems to pinpoint the minimal condition for bypassing Erd\H{o}s' Girth Conjecture. In addition, we also study the sourcewise variant of multiplicative spanners, additive spanners and additive emulators. The main results are summarized below.

\begin{theorem}[Hybrid spanners]
\label{thm:hybrid}
For every integer $k \geq 2$ and unweighted undirected $n$-vertex graph $G=(V,E)$, there exists a (polynomially constructible) subgraph of size $O(k^2 \cdot n^{1+1/k})$ that provides multiplicative stretch $2k-1$ for every pair of neighboring vertices $u$ and $v$ and a multiplicative stretch $k$ for the rest of the pairs. (By Erd\H{o}s' conjecture, providing a multiplicative stretch of $k$ for all the pairs requires $\Omega(n^{1+2/(k+1)})$ edges.)
\end{theorem}

%\textbf{MP: currently do not know how to do it}
%\begin{theorem}[Lower bound for $k$-hybrid spanners]
%\label{thm:lb_hybrid}
%For every integer $k \in [2, O(\log n/\log \log n)]$, there exists an $n$-vertex graph $G=(V,E)$ and such that any $k$-hybrid spanner has at least $\Omega(1/k \cdot n^{1+1/k})$ edges.
%\end{theorem}

\begin{theorem}[Sourcewise spanners]
\label{thm:hybridsw}
For every integer $k \geq 2$, and an unweighted undirected $n$-vertex graph $G=(V,E)$ and for every subset of sources $S \subseteq V$ of size $|S|=O(n^{\varepsilon})$, there exists a (polynomially  constructible) subgraph of size $O(k^2 \cdot n^{1+\varepsilon/k})$ that provides multiplicative stretch $2k-1$ for every pair of neighboring vertices $(u, v) \in S \times V$ and a multiplicative stretch of $2k-2$ for the rest of the pairs in $S \times V$. This subgraph is referred to here as \emph{sourcewise spanner}.
\end{theorem}

\begin{theorem}[Lower bound for additive sourcewise spanners and emulators]
\label{thm:lb}
For every integer $k \in [2, O(\log n/\log \log n)]$ and $\varepsilon \in [0,1]$, there exists an $n$-vertex graph $G=(V,E)$ and a subset of sources $S \subseteq V$ of size $|S|=O(n^{\varepsilon})$ such that any $(2k-1)$-additive sourcewise spanner (i.e., subgraph that maintains a $(2k-1)$-additive approximation for the $S \times V$ distances) has at least $\Omega(n^{1+\varepsilon/k}/k)$ edges. The lower bound holds for additive emulators up to order $O(k)$. For $2$-additive sourcewise emulators there is a matching upper bound.
\end{theorem}

\begin{theorem}[Upper bound for additive sourcewise spanners]
\label{thm:addubadd}
Let $k \geq 1$ be an integer.
(1) For every unweighted undirected $n$-vertex graph $G=(V,E)$ and for every subset of sources $S \subseteq V$, $|S|=O(n^{\varepsilon})$, there exists a (polynomially constructible) $2k$-additive sourcewise spanner with $\widetilde{O}(k \cdot n^{1+(\varepsilon \cdot k+1)/(2k+2)})$ edges.\\
(2) For $|S|=\Omega(n^{2/3})$, there exists a $4$-additive sourcewise spanner with $O(n^{1+\varepsilon/2})$ edges (by the lower bound of Thm. \ref{thm:lb}, any $3$-additive sourcewise spanner requires $\Omega(n^{1+\varepsilon/2})$ edges).
\end{theorem}

The time complexities of all our upper bound construction are obviously polynomial; precise analysis is omitted from this extended abstract.

\subsection{Preliminaries}
We consider the following graph structures.

\dnsparagraph{$(\alpha, \beta)$-spanners}
For a graph $G=(V,E)$, the subgraph $H \subseteq G$ is an $(\alpha, \beta)$-spanner for $G$ if for every $(u,v) \in V \times V$,
\begin{equation}
\label{eq:stand_spanner}
\dist(u, v, H) \leq \alpha \cdot \dist(u, v, G)+\beta~.
\end{equation}
$(\alpha,0)$-spanners (resp., $(1,\beta)$-spanners) are referred to here as $\alpha$-spanners (resp., $\beta$-additive spanners).
\dnsparagraph{Hybrid spanners}
Given a graph $G=(V,E)$, a subgraph $H \subseteq G$ is a $k$-hybrid spanner
iff for every $(u,v) \in V \times V$ it holds that
\begin{equation}
\label{eq:hybrid_spanner}
\dist(u, v,H) \leq
\begin{cases}
(2k-1) \cdot \dist(u, v, G), & \text{if $(u,v) \in E(G)$;}\\
k \cdot \dist(u,v, G), & \text{otherwise.}
\end{cases}
\end{equation}
%Alternatively, for every $(u,v) \in V \times V$, it holds that $\dist(u, v, H)=\max\{\dist(u, v, G)+2(k-1), k \cdot \dist(u, v, G)\}$.
\dnsparagraph{Sourcewise spanners}
Given an unweighted graph $G=(V,E)$ and a subset of vertices $S \subseteq V$, a subgraph $H \subseteq G$ is an \emph{$(\alpha, \beta,S)$-spanner} iff Eq. (\ref{eq:stand_spanner}) is satisfied for every $\langle s, v \rangle \in S \times V$.
When $\beta=0$ (resp., $\alpha=1$), $H$ is denoted by \emph{$(\alpha,S)$-sourcewise spanner} (resp., \emph{$(\beta,S)$-additive sourcewise spanner}).
%A subgraph $H$ is an \emph{$(\alpha, S)$-hybrid sourcewise spanner} iff Eq. (\ref{eq:hybrid_spanner}) is satisfied for every $\langle s, v \rangle \in S \times V$.
\dnsparagraph{Emulators}
Given an unweighted graph $G=(V,E)$, a weighted graph $H=(V, F)$ is an \emph{$(\alpha, \beta)$-emulator} of $G$ iff
$\dist(u, v,G) \leq \dist(u, v,H) \leq
\alpha\cdot\dist(u, v, G)+\beta$ for every $(u, v) \in V \times V$. $(1,\beta)$-emulators are referred to here as $\beta$-additive emulators.
For a given subset of sources $S \subseteq V$, the graph $H=(V,F)$ is a \emph{$(\beta,S)$-additive sourcewise emulator} if the $S \times V$ distances
are bounded in $H$ by an additive stretch of $\beta$.
%
%
%In sourcewise additive and multiplicative emulators the

\subsection{Notation}
For a subgraph $G'=(V', E') \subseteq G$
(where $V' \subseteq V$ and $E' \subseteq E$)
and a pair of vertices $u,v \in V'$, let $\dist(u,v, G')$ denote the shortest-path distance in edges between $u$ and $v$ in $G'$. Let $\Gamma(v,G)=\{u ~\mid~ (u,v) \in E(G)\}$ be the set of neighbors of $v$ in $G$.
For a subgraph $G' \subseteq G$, let $|G'|=|E(G')|$ denote the number of edges in $G'$.
For a path $P=[v_1, \ldots, v_k]$, let $P[v_i, v_j]$ be the subpath of $P$ from $v_i$ to $v_j$. For paths $P_1$ and $P_2$, let $P_1 \circ P_2$ denote the path obtained by concatenating $P_2$ to $P_1$. Let $SP(s, v_i, G')$ be the set of $s-v_i$ shortest-paths in $G'$.
When $G'$ is the input graph $G$, let
$\pi(x,y) \in SP(x,y,G)$ denote some arbitrary $x-y$ shortest path in $G$, hence $|\pi(x,y)|=\dist(x,y,G)$.
For a subset $V' \subseteq V$, let
$\dist(u, V', G)=\min_{u' \in V'}\dist(u,u',G)$. Similarly, for subsets $V_1,V_2 \subseteq V$, $\dist(V_1,V_2,G)=\min_{v_1 \in V_1, v_2 \in V_2} \dist(v_1,v_2,G)$. When the graph $G$ is clear from the context, we may omit it and simply write $\Gamma(u),\dist(u,v), \dist(u,V')$ and $\dist(V_1,V_2)$.
\par A clustering $\mathcal{C}=\{C_1, \ldots, C_{\ell}\}$ is a collection of disjoint subsets of vertices, i.e., $C_i \subseteq V$ for every $C_i \in \mathcal{C}$ and $C_{i} \cap C_{j}=\emptyset$ for every $C_i, C_j \in \mathcal{C}$. Note that a clustering is not necessarily a partition of $V$, i.e., it is not required that $\bigcup_i C_i = V$.
A cluster $C \in \mathcal{C}$ is said to be \emph{connected} in $G$ if the induced graph $G[C]$ is connected.
%For two subsets of vertices $V_1, V_2$, define $\dist(V_1, V_2,G)=\min_{u \in V_1, v \in V_2}\dist(u,v,G)$.
For clusters $C$ and $C'$, let $E(C,C')=(C \times C') \cap E(G)$ be the set of edges between $C$ and $C'$ in $G$. For notational simplicity, let $E(v,C)=E(\{v\},C)$. A vertex $v$ is \emph{incident} to a cluster $C$ if $E(v,C) \neq \emptyset$.
In a similar manner, two clusters $C$ and $C'$ are adjacent to each other if $E(C,C')\neq \emptyset$.
\paragraph{Organization.}
We start with upper bounds. Sec. \ref{sec:hybrid} describes the construction of $k$-hybrid spanners. Sec. \ref{subsec:sourcewise_ub}, presents the construction of $(\alpha, S)$ sourcewise spanners. Then, Sec. \ref{sec:lb_add} presents a lower bound construction for  $(\beta, S)$ sourcewise additive spanners and emulators. Finally, Sec. \ref{sec:up_add} provides an upper bound for $(2k,S)$-additive sourcewise spanners for general values of $k$. In addition, it provides a tight construction for $(2,S)$-additive sourcewise emulators.

\section{Hybrid spanners}
\label{sec:hybrid}
In this section, we establish Thm. \ref{thm:hybrid}.
%and Thm. \ref{thm:lb_hybrid}.
%
%\subsection{Upper Bound}
%\label{subsec:hybrid_ub}
%
For clarity of presentation, we describe a randomized construction whose output spanner has $O(k^2 \cdot n^{1+1/k})$ edges in expectation. Using the techniques of \cite{BSADD10}, this construction can be derandomized with the same bound on the number of edges.
\begin{theorem}
\label{thm:ub_mult}
Let $k \geq 2$ be an integer. For every unweighted $n$-vertex graph $G=(V,E)$ with $m$ edges, a $k$-hybrid spanner $H \subseteq G$ with $O(k^2 \cdot n^{1+1/k})$ edges can be constructed in $O(k^2 \cdot m)$ time.
\end{theorem}
\paragraph{The algorithm.}
We begin by describing a basic procedure $\CLUSTER$, slightly adapted from \cite{BSADD10}, that serves as a building block in our constructions.
For an input unweighted graph $G=(V,E)$, a stretch parameter $k$ and a density parameter $\mu$, Algorithm $\CLUSTER$
iteratively constructs a sequence of $k+1$ clusterings $\mathcal{C}_0, \ldots, \mathcal{C}_{k}$ and a clustering graph $H_{k} \subseteq G$. Each clustering $\mathcal{C}_\tau$ consists of $m_\tau=n^{1-\tau \cdot \mu}$ disjoint subsets of vertices, $C_{\tau}=\{C_1^\tau, \ldots, C_{m_\tau}^\tau\}$. Each cluster $C_j^\tau \in \mathcal{C}_\tau$ is connected and has a \emph{cluster center} $z_j$ satisfying that $\dist(u, z_j, G) \leq \tau$ for every $u\in C_j^\tau$. Denote the set of cluster centers of $\mathcal{C}_\tau$ by $Z_\tau$. These cluster centers correspond to a sequence of samples taken from $V$ with decreasing densities where $V=Z_0 \supseteq Z_1 \supseteq \ldots \supseteq Z_{k}$. On a high level, at each iteration $\tau$, a clustering of radius-$\tau$ clusters is constructed and its shortest-path spanning forest (spanning all the vertices in the clusters), as well as an additional subset of edges $Q_\tau$ adjacent to unclustered vertices, are chosen to be added to the spanner $H_\tau$. We now describe the algorithm $\CLUSTER(G, k, \mu)$ in detail. Assume some ordering on the vertices $V=\{v_1, \ldots, v_n\}$. Initially, the cluster centers are $Z_0=V=\{v_1, \ldots, v_n\}$, where each vertex forms its own cluster of radius $0$, hence $\mathcal{C}_0=\{\{v\} ~\mid v \in V\}$ and the spanner is initiated to $H_0=\emptyset$. At iteration $\tau \geq 1$, a clustering $\mathcal{C}_\tau$ is defined based on the cluster centers $Z_{\tau-1}$ of the previous iteration. Let $Z_{\tau} \subseteq Z_{\tau-1}$ be a sample of $m_\tau=O(n^{1-\tau \cdot \mu})$ vertices chosen uniformly at random from $Z_{\tau-1}$. The clustering $\mathcal{C}_\tau$ is obtained by assigning every vertex $u$ that satisfies $\dist(u, Z_{\tau}, G) \leq \tau$ to its closest cluster center $z \in Z_\tau$, i.e., such that $\dist(u,z, G)=\dist(u, Z_\tau,G)$. If there are several cluster centers in $Z_\tau$ at distance $\dist(u, Z_\tau,G)$ from $u$, then the closest center with the minimal index is chosen.

Formally, for a vertex $v$ and subset of vertices $B$, let $\nearest(u,B)$ be the closest vertex to $u$ in $B$ where ties are determined by the indices, i.e., letting $B'=\{u_1,\ldots, u_\ell\} \subseteq B$ be the set of closest vertices to $u$ in $B$, namely, satisfying that $\dist(v,u_1)=...=\dist(v,u_\ell)=\dist(v,B)$, then $\nearest(u,B) \in B'$ and has the minimal index in $B'$.
%
%let $v_{\tau_1}, \ldots, v_{\tau_\ell}\in Z_\tau$ be such that $\dist(u, v_{\tau_j},G)=\dist(u, Z_\tau, G)$ for every $j \in \{1, \ldots, \ell\}$ and $\tau_1 < \tau_2 < \ldots< \tau_\ell$.
Then $u$ is assigned to the cluster of the center $\nearest(u,Z_\tau)$.
Add to $H_\tau$ the forest $F_\tau$ consisting of the radius-$\tau$ spanning tree of each $C \in \mathcal{C}_\tau$.
Note that the  definition of the clusters immediately implies their connectivity.
%\begin{observation}
%\label{obs:clust_connect}
%Every $C \in \mathcal{C}_\tau$ is connected in $H_\tau$.
%\end{observation}
Next, an edge set $Q_\tau$ adjacent to unclustered vertices is added to $H_\tau$ as follows.
Let $\Delta_\tau$ denote the set of vertices
that occur in each of the clusterings $\mathcal{C}_0, \ldots, \mathcal{C}_{\tau-1}$ but do not occur in $\mathcal{C}_{\tau}$.
(Observe that such a vertex may re-appear again in some future clusterings.)
Formally, let $\widehat{V}_\tau=\bigcup_{C \in \mathcal{C}_\tau} C$ be the set of vertices that occur in some cluster in the clustering $\mathcal{C}_\tau$. Then,
$\Delta_\tau=\left(\bigcap_{j=0}^{\tau-1} \widehat{V}_j \right) \setminus \widehat{V}_\tau~$. Note that by this definition, each vertex belongs to at most one set $\Delta_\tau$. Hence we have:
\begin{observation}
\label{obs:delta_disjoint}
The sets $\Delta_\tau$ are disjoint.
\end{observation}
For every vertex $v \in \Delta_\tau$ and every cluster $C \in \mathcal{C}_{\tau-1}$ that is adjacent to $v$, pick one vertex $u \in C$ adjacent to $v$ and add the edge $(u,v)$ to $Q_\tau$. (In other words, an edge $(u,v)$ is \emph{not} added to $Q_\tau$ for $v \in \Delta_{\tau}$ if either $u \notin \widehat{V}_{\tau-1}$ or an edge $(u',v)$ was added to $Q_\tau$ where $u'$ and $u$ are in the same cluster $C \in \mathcal{C}_{\tau-1}$.) Then add $Q_\tau$ to $H_\tau$.
This completes the description of Algorithm $\CLUSTER$; a pseudocode is given below.
\dnsparagraph{Algorithm $\CLUSTER(G,k,\mu)$}
\begin{description}
\item{(T1)}
Let $H_0=\emptyset$ and $Z_0=n$. Select a sample $Z_\tau$ uniformly at random from $Z_{\tau-1}$ with probability $n^{-\mu}$ for $\tau=1$ to $k$ (if $\mu=1$ and $\tau=k$, set $Z_{k}=\emptyset$).
\item{(T2)}
For $\tau=1$ to $k$, define the clustering $\mathcal{C}_\tau$ by adding the $\tau$-radius neighborhood for all cluster centers $Z_\tau$, i.e., every $u \in V$ satisfying $\dist(u, Z_\tau) \leq \tau$ is connected to $\nearest(u, Z_\tau)$.
Let $F_\tau$ denote the $\tau$-radius neighborhood forest corresponding to $\mathcal{C}_\tau$.
\item{(T3)}
For every vertex $v \in \Delta_\tau$ that was unclustered in the clustering $\mathcal{C}_{\tau}$ for the first time, let $e(v,C)$ be an arbitrary edge from $E(v, C)$ for every $C \in \mathcal{C}_{\tau-1}$.
\item{(T4)}
$H_{\tau}=H_{\tau-1} \cup F_{\tau} \cup \{e(v,C) ~\mid~ v \in \Delta_\tau, C \in \mathcal{C}_{\tau-1}\}$.
\end{description}
The first step of Algorithm $\HYBRID$ applies Algorithm $\CLUSTER(G, k, \mu)$ for $\mu=1/k$, resulting in the graph $H_{k}$.
Note that by Thm. 3.1 of \cite{BSADD10}, $H_{k}$ is a $(2k-1)$ spanner. Hence, the stretch for neighboring vertices is $(2k-1)$ as required. We now add two edge sets to $H_{k}$ in order to provide a multiplicative stretch $k$ for the remaining pairs.
Let
\begin{equation}
\label{eq:t}
t=\lfloor k/2 \rfloor \mbox{~~and~~} t'=k-1-t,
\end{equation}
Note that $t'=t$ when $k$ is odd and $t'=t-1$ when $k$ is even, so in general $t' \leq t$.
\par
\begin{figure}[htb!]
\begin{center}
\includegraphics[scale=0.5]{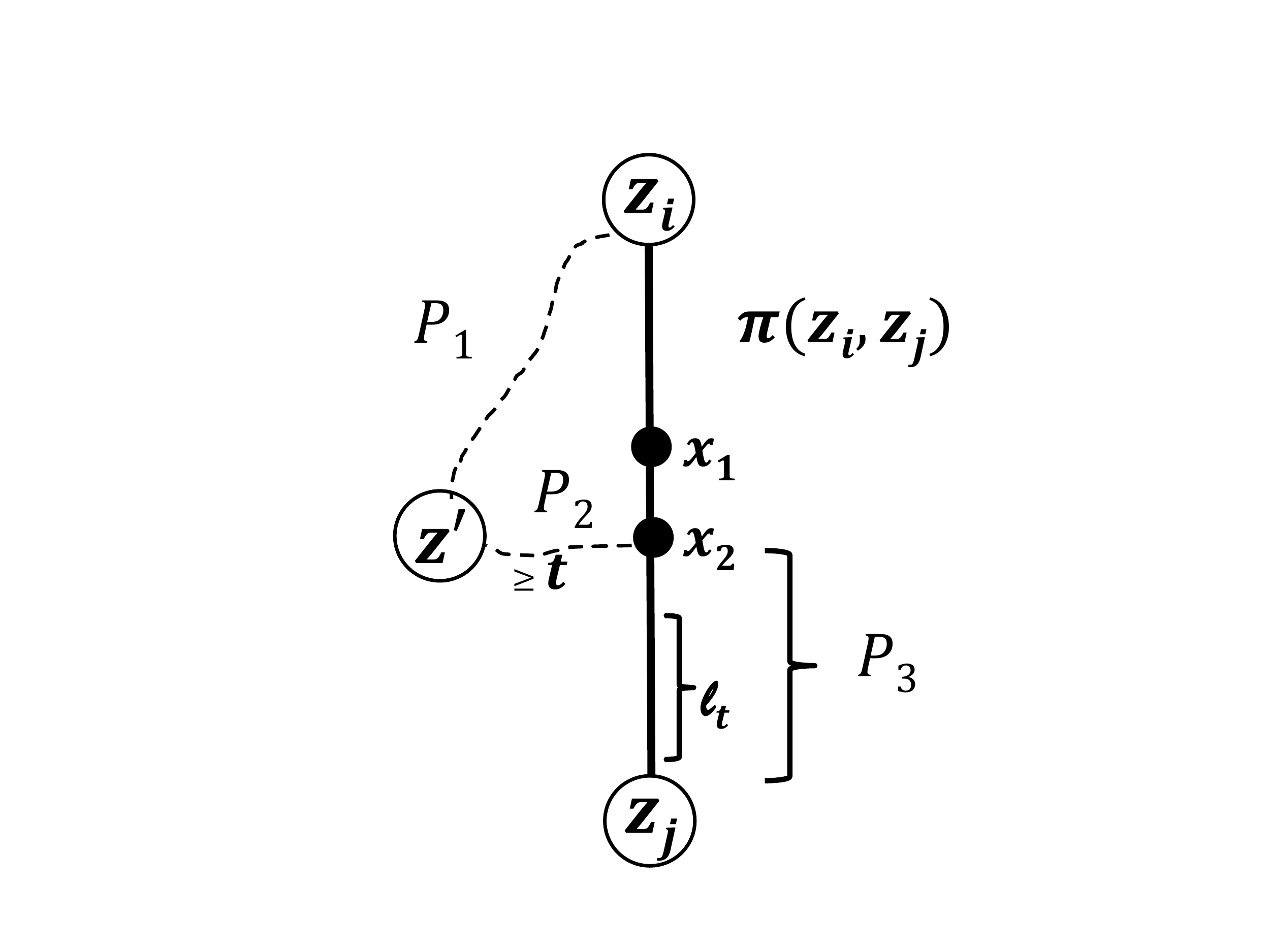}
\caption{\label{fig:clustercenters}
Illustration of case (b) for Lemma \ref{cl:tclustcenters}(2).
If not all edges on $\pi(z_i, z_j)$ are $t$-clustered, then the induction assumption for cluster centers preceding $z_j$ in the ordering can be used; this is due to the fact that the last $\splenone$ edges of $\pi(z_i, z_j)$ were taken into $H$. The dashed lines represent bypasses in $H$.}
\end{center}
\end{figure}

\begin{figure}[htb!]
\begin{center}
\includegraphics[scale=0.4]{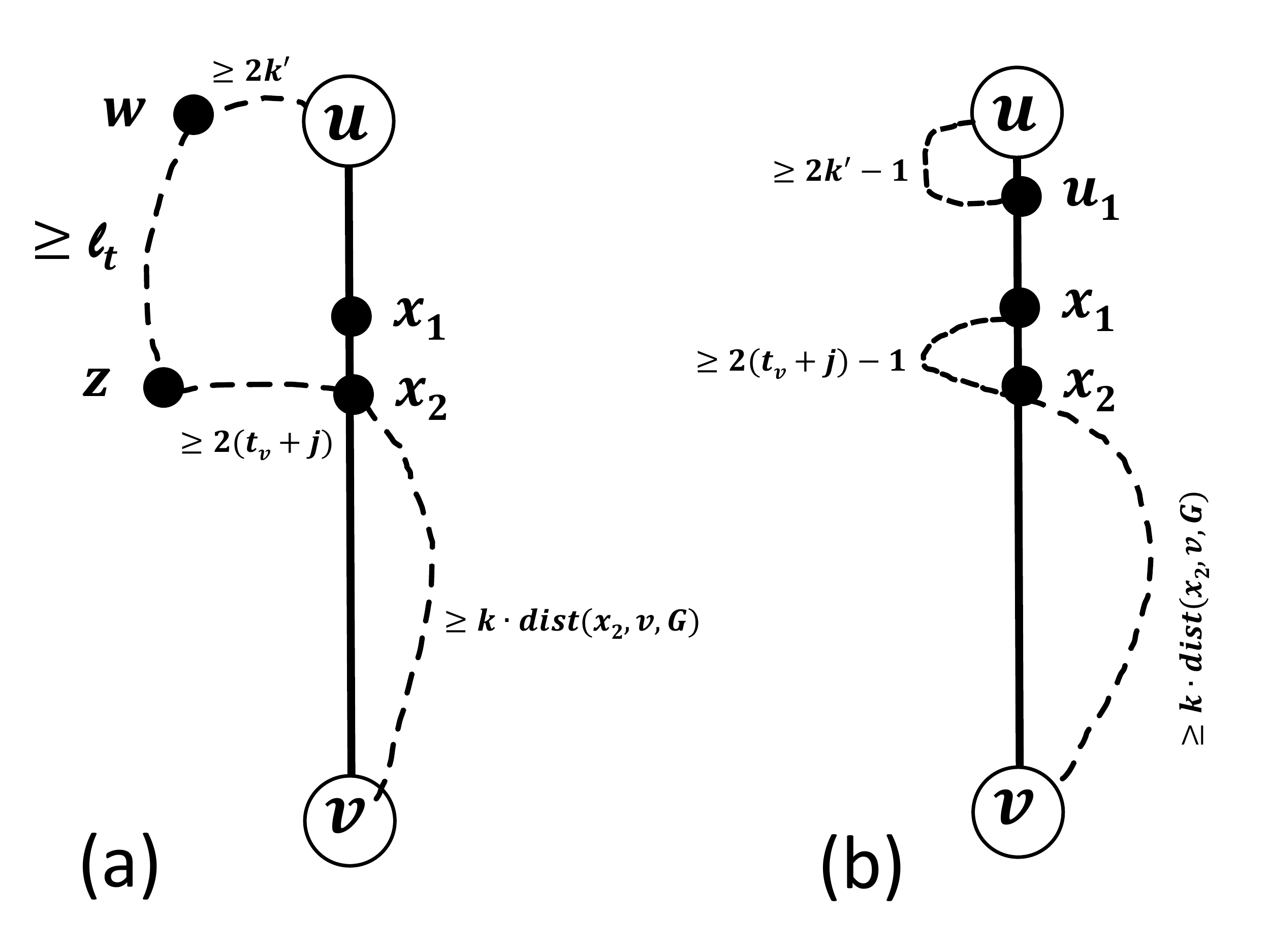}
\caption{\label{fig:allpairs}
Illustration of Case (a) for Lemma \ref{lem:spann_correct}.
The dashed lines represent bypasses in $H$.
(a) Case (a1). Since $u$ is $k'$-clustered and $x_2$ is $(t_v+j)$-clustered and in addition, $\dist(u, v, G) \leq \splentwo$ it holds that the shortest path between the corresponding clusters has been added to $H$.
(b) Case (a2). Since $u$ is $k'$-unclustered, its immediate edges, specifically, $(u_0, u_1)$ has small stretch in $H$ that compensates for the extra large stretch of the edge $(x_1, x_2)$ in $H$.}
\end{center}
\end{figure}

 The algorithm considers the collection of $Z_{t'} \times Z_t$ shortest paths $\mathcal{P}=\{\pi(z_i, z_j) \mid z_i \in Z_{t'} \mbox{~and~} z_j \in Z_t\}$. Starting with $H=H_{k}$, for each path $\pi(z_i, z_j) \in \mathcal{P}$, it adds to $H$ the $\splenone$ last edges of $\pi(z_i, z_j)$ (closest to $z_i$), where
\begin{equation}
\label{eq:ellk}
\splenone=7t+8t^2~.
\end{equation}
For every pair of clusters $C_1,C_2$, let $\pi(C_1,C_2)$ denote the shortest path in $G$ between some closest vertices $u_1\in C_1$ and $u_2 \in C_2$ (i.e., $\dist(C_1,C_2,G)=\dist(u_1,u_2,G)$).
For every $\tau$ from $0$ to $k-1$, and for every pair $C_1 \in \mathcal{C}_{\tau}$ and $C_2 \in \mathcal{C}_{k-1-\tau}$, the algorithm adds to $H$, the $\ell$ last edges of $\pi(C_1,C_2)$, where $\ell=\splenone$ for $\tau \in \{t',t\}$ and $\ell=2k-1$ otherwise.
This completes the description of Algorithm $\HYBRID$, whose summary is given below.
\dnsparagraph{Algorithm $\HYBRID$}
\begin{description}
\item{(S1)}
Let $H_{k}=\CLUSTER(G, k, 1/k)$.
\item{(S2)}
Let $E_2$ be the edge set containing the last $\splenone$ edges of the path $\pi(z_i,z_j)$ for every $z_i \in Z_{t'}$ and $z_j \in Z_t$.
\item{(S3)}
Let $E_3$ be the edges set containing, for every $\tau \in \{0, \ldots, k-1\}$, and for every $C_1 \in \mathcal{C}_{\tau}$ and $C_2 \in \mathcal{C}_{k-1-\tau}$, the last $\ell$ edges of the path $\pi(C_1,C_2)$ where $\ell=\splenone$ for $\tau \in \{t',t\}$ and $\ell=2k-1$ otherwise.
%for the closest vertices $u \in C_1$ and $v \in C_2$ (i.e., $u$ and $v$ is  an arbitrary pair in $C_1 \times C_2$ satisfying that $\dist(u,v)=\dist(C_1,C_2)$).
\item{(S4)}
Let $H \gets H_{k} \cup E_2 \cup E_3$.
\end{description}
%In Appendix Sec. \ref{append:hybrid},
Note that our algorithm bares some similarity to the $(k,k-1)$ construction of \cite{BSADD10} but the analysis is different. The key difference between these two constructions is that in \cite{BSADD10} only edges (i.e., shortest-path of length $1$) are added between certain pairs of clusters. In contrast, in our construction, $O(k^2)$ edges are taken from each shortest-path connecting the close-most vertices coming from certain subset of clusters. This allows us to employ an inductive argument on the desired \emph{purely} multiplicative stretch, without introducing an additional additive stretch term (for non-neighboring pairs) as in the $(k,k-1)$ construction of \cite{BSADD10}. Specifically, by adding paths of length $\ell_t$ between center pairs in $Z_{t'} \times Z_t$, a much better stretch guarantee can be provided for (non-neighboring) $Z_{t'} \times Z_t$ pairs: a multiplicative stretch $k$ plus a \emph{negative} additive term. This additive term is then increased but in a controlled manner (due to step (S3)), resulting in a \emph{zero} additive term for \emph{any} non-neighboring vertex pair in $V \times V$.
%(i.e., by taking  nearby cluster pairs of distance at most $O(k^2)$ as the induction base).
%are guaranteed to have good stretch (in fact no stretch) in the resulting spanner, hence establishing the induction base.
\paragraph{Analysis}
We begin with size analysis. An edge $e \in E(G)$ is \emph{missing} if it is not included in the spanner, i.e., $e \in E(G) \setminus E(H)$. We first bound the expected size of the partial spanner $H_{k}$ and the final spanner $H$ obtained at the end of phase (S1).
\begin{lemma}
\label{lem:sspanner}
(1) $\mathbb{E}(|H_{k}|)=O(n^{1+\mu}+k \cdot n)$.\\
(2) $\mathbb{E}(|H|)=O(k^2 \cdot n^{1+1/k})$.
\end{lemma}
\Proof
%Begin with (1). The number of edges contributed by the radius-$\tau$ spanning trees of the forest $F_\tau$ added in the $\tau$'th iteration is at most $n$ , summing up to at most $kn$ over the $k$ iterations.
%It remains to bound the number of edges added in the sets $Q_\tau$. We claim that the expected number of edges in the set $Q_\tau$ contributed by the vertices in $\Delta_\tau$ is $\mathbb{E}(Q_\tau) \leq  |\Delta_\tau| \cdot n^{\mu}$.
%Consider a vertex $v \in \bigcup_{j=0}^{\tau-1} \widehat{V}_j$ (i.e., that appears in \emph{each} of the clustering $\mathcal{C}_0, \ldots, \mathcal{C}_{\tau-1}$) and let $s$ be the number of clusters in $\mathcal{C}_{\tau-1}$ that are adjacent to $v$.
%The probability for $v$ to be included in $\mathcal{C}_\tau$ is at least $1-(1-n^{-\mu})^s$. For each of these $s$ clusters, exactly one edge is added to $H_\tau$ due to $v$ (it might be that additional edges would be added to $H_\tau$, but this would be due to their other endpoint of the edge), hence the expected number of edges contributed by $v$ is at most $s \cdot \mathbb{P}[v\mbox{~ does not appear in ~}\mathcal{C}_\tau] \leq s \cdot (1-n^{-\mu})^s \leq n^{\mu}$, and $\mathbb{E}(Q_\tau) \leq |\Delta_\tau| \cdot n^{\mu}$. By Obs. \ref{obs:delta_disjoint}, $\mathbb{E}(|\bigcup_\tau Q_\tau|)\leq \sum_\tau |\Delta_\tau|\cdot n^{\mu}\leq n\cdot n^{\mu}$.
Part (1) follows by the same argumentation as in Lemma 2.9 of \cite{BSADD10}. Consider Part (2). Since $\mu=1/k$, it holds that $H_{k}$ obtained at the end  of phase (S1) contains $O(n^{1+1/k})$ edges. We now bound the number of edges added in phase (S2), $|E_2|=\splenone \cdot |Z_{t}| \cdot |Z_{k-1-t}|$, hence $|E_2|=\splenone \cdot n^{1-\tau/k} \cdot n^{1-(k-1-\tau)/k} =\splenone \cdot n^{1+1/k}$. Finally, in phase (S3), $(2k-1) \cdot|Z_{\tau}| \cdot |Z_{k-1-\tau}|$ edges are added for every $\tau \in \{0, \ldots, k-1\} \setminus \{t,t'\}$ and in addition, $\splenone \cdot n^{1-t/k} \cdot n^{1-t'/k}$ edges connecting closest pairs in the clusters of $\mathcal{C}_t$ and $\mathcal{C}_{t'}$ are added as well. Since $|Z_{\tau}|= n^{1-\tau/k}$ and $|Z_{k-1-\tau}|= n^{1-(k-1-\tau)/k}$, it holds that $|E_3|=O(k^2 \cdot n^{1+1/k})$.
\QED
%
%\begin{observation}
%\label{obs:multspanner_size_final}
%$\mathbb{E}(|H|)=O(k^2 \cdot n^{1+1/k})$.
%\end{observation}
%\def\APPENDNEDGE{
%\Proof
%Since $\mu=1/k$, it holds that $H_{k-1}$ obtained at the end  of phase (S1) contains $O(n^{1+1/k})$ edges. We now bound the number of edges added in phase (S2), $|E_2|=|Z_{t}| \cdot |Z_{k-1-t}| \cdot \splenone$. Since $|Z_{\tau}|= n^{1-\tau/k}$ and $|Z_{k-1-\tau}|= n^{1-(k-1-\tau)/k}$, it holds that $|E_2|=\splenone \cdot n^{1-\tau/k} \cdot n^{1-(k-1-\tau)/k} =\splenone \cdot n^{1+1/k}$.
%Finally, in phase (S3), $|E_3|=|Z_{\tau}| \cdot |Z_{k-1-\tau}| \cdot \splentwo$ edges are added. It holds that $|E_3|=\splentwo \cdot n^{1+1/k}$. By plugging Eq. (\ref{eq:ellk}), the claim follows.
%\QED
We now turn to establish correctness. The following notation is useful in our analysis. A vertex is $\ell$-clustered if it belongs to some $C \in \mathcal{C}_\ell$. For an $\ell$-clustered vertex $v$, let $z_\ell(v) \in Z_\ell$ denote the cluster center of $v$ in $\mathcal{C}_\ell$.
An edge $e=(u,v)$ is $\ell$-clustered if both $u$ and $v$ are $\ell$-clustered, otherwise it is $\ell$-unclustered. The next lemma (see Thm. 3.1 in \cite{BSADD10}) plays a major role in our stretch analysis.
\begin{lemma}
\label{cl:clut_edge}
For every $\ell$-unclustered edge $(u,v)$, it holds that $\dist(u, v, H) \leq 2\ell-1$.
\end{lemma}
\Proof
Let $\ell'$ be the minimum index such that either $u$ or $v$ was unclustered in $\mathcal{C}_{\ell'}$ (clearly, $\ell' \leq \ell$) and without loss of generality let the unclustered vertex be $u$.
Then by Phase (T3), there is an edge $(u,w)$ in $H_{\ell}$ from $u$ to some vertex in $C_{\ell'-1}(v)$. In addition, there is a path in $H_{\ell}$ from $w$ to $v$ of length at most $2(\ell'-1)$, twice the radius of $C_{\ell'-1}(v)$. Since $\ell' \leq \ell$, it follows that $\dist(u, v, H) \leq 2\ell-1$.
\QED
We begin by considering the stretch between pairs of cluster centers $Z_{t'} \times Z_t$.
\begin{lemma}
\label{cl:tclustcenters}
For every pair of cluster centers $z_i, z_j \in Z_{t'} \times Z_t$ it holds that\\
(1) If $\dist(z_i, z_j,G) \leq \splenone$, then $\dist(z_i, z_j, H)=\dist(z_i, z_j,G)$.\\
(2) If $\dist(z_i, z_j,G)>\splenone$, then $\dist(z_i, z_j, H) \leq 2t \cdot \left(\dist(z_i, z_j,G)+1 \right)-\splenone$.
%If $\dist(z_i, z_j,G)>\splenone$, then $\dist(z_i, z_j, H) \leq 2t \cdot \dist(z_i, z_j,G)+1)-\splenone$.
\end{lemma}
\Proof
Fix a cluster center $z_i \in Z_{t'}$ and let the cluster centers $Z_{t}=\{z_1, \ldots, z_{\ell}\}$ be ordered in nondecreasing distance from $z_i$, i.e., $\dist(z_i, z_1, G) \leq \dist(z_i, z_2, G) \leq \ldots \leq \dist(z_i, z_{\ell},G)$.
Assume, towards contradiction, that the the claims do not hold, and let $z_j$ be the first center in the ordering for which one of the claims does not hold. Define $D_G=\dist(z_i, z_j, G)$ and $D_H=\dist(z_i, z_j, H)$. Since the last $\splenone$ edges of the path $\pi(z_i, z_j)$ are taken into $H$, it holds that $z_j$ does not satisfy claim (2) and hence
\begin{eqnarray}
\label{eq:dg}
D_G=|\pi(z_i, z_j)| > \splenone~.
\end{eqnarray}
We now distinguish between cases depending on the type of the edges missing from $\pi(z_i, z_j)$ in $H$.\\
Case (a): all missing edges in $\pi(z_i, z_j)\setminus H$ are $t$-unclustered. By Lemma \ref{cl:clut_edge} and Eq. (\ref{eq:dg}), we have that
$D_H \leq (2t-1)D_G < 2t \cdot D_G -\splenone$.\\
Next, consider the complementary case (b): some of the missing edges of $\pi(z_i, z_j) \setminus H$ are $t$-clustered.
Let $e=(x_1, x_2)$ be the last missing edge on the path $\pi(z_i, z_j)$ (the edge closest to $z_j$) which is $t$-clustered, hence $x_2$ is $t$-clustered.  Let $z'=z_{t}(x_2)$ be the cluster center of $x_2$. For an illustration, see Fig. \ref{fig:clustercenters}.
We first claim that $z'$ precedes $z_j$ in the ordering.  To see this, recall that the last $\splenone$ edges of the path $\pi(z_i, z_j)$ were taken into $H$ , thus $\dist(z_i, x_2, G) \leq \dist(z_i, z_j, G)-\splenone=D_G-\splenone$.
Since the radius of each cluster in $\mathcal{C}_t$ is at most $t$, we get that
\begin{equation}
\label{eq:clustcloser}
\dist(z_i, z', G) \leq \dist(z_i, x_2, G)+t \leq D_G-\splenone+t<D_G~,
\end{equation}
where the strict inequality follows by Eq. (\ref{eq:ellk}).
This strict inequality implies that indeed $z'$ precedes $z_j$ in the ordering and by the definition of $z_j$ it follows that $z'$ satisfies the lemma.
Consider the alternative $z_i-z_j$ path $P=P_1 \circ P_2 \circ P_3$ where $P_1 \in SP(z_i, z',H)$, $P_2 \in SP(z', x_2,H)$ and $P_3 \in SP(x_2, z_j, H)$.
Since $P \subseteq H$, it remains to bound its length.
Since $z'$ is the cluster center of $x_2$, $|P_2| \leq t$. Since $P_3$ is free from missing $t$-clustered edges, by Lemma \ref{cl:clut_edge}, $|P_3| \leq (2t-1)\cdot \dist(x_2, z_j,G)$. To bound the length of $P_1$ we distinguish between two subcases depending on the length of $\pi(z_i, z')$. Subcase (b1): $|\pi(z_i, z')|\leq \splenone$. Then, $\dist(z_i, z', H)=\dist(z_i, z', G)$ and overall we have that
\begin{eqnarray*}
D_H \leq |P|&=&|P_1|+|P_2|+|P_3| ~\leq~ \dist(z_i, z', G)+t+(2t-1)\dist(x_2, z_j, G)
\\&\leq&
D_G-\splenone+2t+(2t-1)\cdot \dist(x_2, z_j, G) ~\leq~ 2t \cdot (D_G+1)-\splenone~,
\end{eqnarray*}
where the second inequality follows by the second inequality of Eq. (\ref{eq:clustcloser}). This contradicts the fact that $z_j$ violates the claim.\\
Subcase (b2): $\dist(z_i, z', G) \geq \splenone+1$. Since $z'$ satisfies the lemma,
\begin{eqnarray}
\label{eq:clusterok}
\dist(z_i, z', H) \leq 2t(\dist(z_i, z', G)+1)-\splenone~.
\end{eqnarray}
Overall, we get that
\begin{eqnarray}
D_H &\leq& |P|=|P_1|+|P_2|+|P_3|  \nonumber
\\&\leq&
2t(\dist(z_i, z', G)+1)-\splenone+t+(2t-1) \cdot \dist(x_2, z_j,G) \label{eq:tmd1}
\\&\leq&
\dist(z_i, z', G)+(2t-1)\cdot\dist(z_i, z', G)+3t-\splenone+(2t-1) \cdot \dist(x_2, z_j,G) \nonumber
\\&\leq&
D_G-\splenone+4t+(2t-1)\cdot \dist(z_i, z', G)-
\splenone+(2t-1) \cdot \dist(x_2, z_j,G) \label{eq:tmd3}
\\&\leq&
D_G-\splenone+4t+(2t-1)\cdot (\dist(z_i, x_2, G)+t)-
\splenone+(2t-1) \cdot \dist(x_2, z_j,G) \nonumber%\label{eq:tmd4}
\\&\leq&
2t\cdot D_G -\splenone,
\end{eqnarray}
where Eq. (\ref{eq:tmd1}) follows by Eq. (\ref{eq:clusterok}), Ineq. (\ref{eq:tmd3})
follows by the  inequality of Eq. (\ref{eq:clustcloser}) and the penultimate inequality follows by the fact that $z'$ and $x_2$ belong to the same radius-$t$ cluster.
In contradiction to the definition of $z_j$. Claim (2) follows.
\QED
Let $V_i$ be the set of $i$-clustered vertices.
We now turn to bound the stretch between pairs of vertices in $V_{t'} \times V_t$.
\begin{lemma}
\label{cl:tclusvertices}
For every non-neighboring vertex pair $x_1\in V_{t'}$ and $x_2  \in V_t$ (i.e., such that $\dist(x_1, x_2,G)\geq 2$),
$\dist(x_1, x_2, H)\leq k\cdot \dist(x_1, x_2,G)$.
\end{lemma}
\Proof
%Begin with Claim (a). Since $\dist(x_1,x_2,G)=1$, it holds that also $\dist(C_{t'}(x_1), C_t(x_2),G)\leq 1$ and therefore by phase (S3) also $\dist(C_{t'}(x_1), C_t(x_2),H) \leq 1$. That is, either there exists a vertex $w \in C_{t'}(x_1) \cup C_t(x_2)$ or an edge $(w,z)$ such that $w \in C_{t'}(x_1)$ and $z \in C_t(x_2)$ was added to $H$ and by using the intercluster connections it holds that
%$\dist(x_1, x_2,H) \leq \dist(x_1, w, H)+1+\dist(z, x_2, H)=
%2t'+1+2t \leq 2k-1$ as required.
%Consider Claim (b) of the lemma where $\dist(x_1, x_2,G)\geq 2$. 
We consider the following cases.\\
Case (1): $\dist(C_{t'}(x_1), C_{t}(x_2), G) \leq \splenone$.
Then again by phase (S3), there exists an $w-z$ path $P$ in $H$ such that $w \in C_{t'}(x_1)$ and $z \in C_{t}(x_2)$ and $|P|=\dist(w, z, G)\leq \dist(x_1, x_2,G)$.
Hence
\begin{eqnarray}
\dist(x_1, x_2, H) &\leq& \dist(x_1, w, H)+\dist(w, z, H)+\dist(z, x_2, H) \nonumber
\\&=&
\dist(x_1, w, H)+\dist(w, z, G)+\dist(z, x_2, H) \nonumber
\\&\leq &
2t'+\dist(w, z, G)+2t \leq 2(k-1)+\dist(x_1, x_2, G) \label{eq:intert1}
\\&\leq &
k\cdot \dist(x_1, x_2, H)~ \nonumber,
\end{eqnarray}
where Eq. (\ref{eq:intert1}) follows by the fact that $x_1$ (resp., $x_2$) and $w$ (resp., $z$) belongs to the same cluster of diameter $2t'$ (resp., $2t$), and $w$ and $z$ are the closest pair in $C_{t'}(x_1)$ and $C_{t}(x_2)$. Finally, the last inequality follows as $\dist(x_1, x_2, H)\geq 2$.\\
Case (2): $\dist(C_{t'}(x_1), C_{t}(x_2), G)> \splenone$.
Let $z_1=z_{t'}(x_1)$ (resp., $z_2=z_t(x_2)$)  be the cluster center of $x_1$ (resp., $x_2$) in the clustering $\mathcal{C}_{t'}$ (resp., $\mathcal{C}_{t}$).
Since $\dist(z_1, z_2,G)\geq \dist(C_{t'}(x_1), C_{t}(x_2), G)>\splenone$, by Lemma \ref{cl:tclustcenters}, it holds that \\ $\dist(z_1, z_2, H) \leq 2t \cdot (\dist(z_1, z_2,G)+1)-\splenone$.
Hence,
\begin{eqnarray}
\dist(x_1, x_2, H) &\leq& \dist(x_1, z_1, H)+\dist(z_1, z_2, H)+\dist(z_2, x_2, H) \nonumber
\\&\leq&
2t+\dist(z_1, z_2, H)+2t' \label{eq:incluine1}
\\&\leq&
2(k-1)+2t\cdot \dist(z_1, z_2,G)+2t -\splenone \label{eq:incluine2}
\\&\leq&
2(k-1)+2t \cdot (\dist(x_1, x_2,G)+2(k-1))-\splenone \label{eq:incluine3}
\\&\leq&
k \cdot \dist(x_1, x_2,G)~, \nonumber
\end{eqnarray}
where Eq. (\ref{eq:incluine1}) follows by the intercluster connections in clusters of diameter $2t$ and $2t'$, Eq. (\ref{eq:incluine2}) follows by Eq. (\ref{eq:t}) and by Cl. \ref{cl:tclustcenters}, Eq. (\ref{eq:incluine3}) follows by the triangle inequality using the fact that $z_1$ and $x_1$ are in the same cluster of diameter $2t'$ and $z_2$ and $x_2$ are in the same cluster of diameter $2t$.
The claim follows.
\QED
Finally, we are ready to to bound the distance for every $u, v \in V \times V$. We have the following.
\begin{lemma}
\label{lem:spann_correct}
For every $(u, v)\in V \times V$:\\
(1) $\dist(u,v, H)\leq 2k-1$ if $(u, v) \in E$.\\
(2) $\dist(u,v, H)\leq k\cdot \dist(u, v,G)$, otherwise.
\end{lemma}
\Proof
Part (1) follows by the fact that $H_{k}$ (the resulting subgraph of step (S1)) is a $(2k-1)$ spanner (see Thm. 3.1 of \cite{BSADD10}). Consider part (2) and let $\pi(u,v)$ be the $u-v$ shortest path in $G$.
Clearly, if $u$ is $t'$-clustered and $v$ is $t$-clustered or vice versa then the lemma follows by Lemma \ref{cl:tclusvertices}. Hence, it remains to consider the complementary case. From now on, assume without loss of generality that $u$ is $t_u$-\emph{unclustered} for a fixed $t_u\in \{t, t'\}$ and define $t_v=k-1-t_u$ ($v$ can be either $t_v$-clustered or $t_v$-unclustered).
%\par Consider the $u-v$ shortest path $\pi(u,v)$.
If all the edges in $\pi(u,v)$ are $t$-unclustered
then by Lemma \ref{cl:clut_edge}, $\dist(u, v, H) \leq (2t-1) \cdot \dist(u, v,G)$ and we are done. Hence, from now on, assume that $\pi(u,v)$ contains at least two $t$-clustered vertices.
In the same manner, if all the edges in $\pi(u,v)$ are $t'$-unclustered then by Lemma \ref{cl:clut_edge}, $\dist(u, v, H) \leq (2t'-1) \cdot \dist(u, v,G)$ and we are done. Therefore $\pi(u,v)$ contains at least two $t_u$-clustered vertices and at least two $t_v$-clustered vertices.
Let $x_1$ be the closest $t_u$-clustered vertex to $u$ and let $x_2$ be the closest $t_v$-clustered vertex to $v$.
%If $|\pi(u,v)|=1$, then since $u$ is $t_u$ unclustered, it holds by Lemma \ref{cl:clut_edge} that $\dist(u, v, H) \leq 2 \cdot t_u-1 \leq k-1$ and we are done.
%Hence, from now on consider the case where $|\pi(u,v)|\geq 2$.
Note that by definition, $u \neq x_1$.
%By the assumption that there are at least two $t$-clustered  vertices and at least two $t'$-clustered vertices, it holds that $x_1 \neq x_2$ and in addition, since $u$ is $t_u$-unclustered it holds that $u \neq x_1 \neq x_2$ and also that $|\pi(u,v)|\geq 2$ (since the case where $|\pi(u,v)|=1$ implies that $u=x_1$ and $v=x_2$).
To prove Claim (2) of the lemma, we partition the path $\pi(u,v)$ into three segments according to $x_1$ and $x_2$. Let $\pi_1=\pi(u, x_1)$, $\pi_2=\pi(x_1, x_2)$ and $\pi_3=\pi(x_2, v)$.
By Lemma \ref{cl:clut_edge}, and recalling that $t' \leq t$ it holds that
\begin{eqnarray}
\label{eq:boundpi}
\dist(u, x_1, H) &\leq& (2t-1) \cdot \dist(u, x_1, G)
\\&\mbox{~and~}&
\dist(x_2, v, H) \leq (2t-1) \cdot \dist(x_2, v, G). \nonumber
\end{eqnarray}
First consider the case where $\pi_1 \circ \pi_2 \subseteq \pi(u,v)$ (e.g., in such a case either $x_1=x_2$ or $x_1$ is strictly closer to $v$ than $x_2$). By Eq. (\ref{eq:boundpi}), it holds that every edge $(y_1,y_2) \in \pi(u,v)$ is either $t$-unclustered or $t'$-unclustered and hence $\dist(y_1,y_2,H) \leq 2t-1$, concluding that $\dist(u,v,H)\leq k \cdot \dist(u,v,G)$ as desired.
From now on, we therefore consider the case where $x_1$ appears on $\pi(u,v)$ \emph{before} $x_2$ (i.e., closer to $u$) and thus $u \neq x_1 \neq x_2$.
We now consider the following case analysis.
\paragraph{Case (a) $\dist(x_1, x_2,G)=1$ and $\dist(u, x_2,G)\leq 2k-1$.}
Let $j \in \{0, \ldots, k-1-t_v\}$ be the maximum number satisfying that $x_2$ is $(t_v+j)$-clustered.
Since $x_2$ is $t_v$-clustered, there exists an $j$ in this range. We now consider two subcases; for a schematic illustration see Fig. \ref{fig:allpairs}.
\paragraph{Case (a1): $u$ is $(k-1-(t_v+j))$-clustered.}
Let $k'=(k-1-(t_v+j))$.
Since $\dist(u, x_2,G)\leq 2k$, it also holds that $\dist(C_{k'}(u), C_{t_v+j}(x_2),G)\leq 2k$, hence by Phase (S3), there exists a $w-z$ path $P$ in $H$ where $w \in C_{k'}(u)$ and $z \in C_{t_v+j}(x_2)$ and $|P|=\dist(w, z, G) \leq \dist(u, x_2, G)$. Recall that as $|P| \leq 2k$, it was taken entirely into $H$. We therefore have the following.
\begin{eqnarray}
\dist(u, v, H) &=& \dist(u, x_2, H)+\dist(x_2, v, H) \nonumber
\\&\leq&
\dist(u, w, H)+\dist(w, z, H)+\dist(z, x_2,H) \nonumber
\\&+&(2t-1) \cdot  \dist(x_2, v, G) \label{eq:tmdd1}
\\&\leq&
2k'+\dist(w, z, G)+2(t_v+j) +(2t-1)\cdot  \dist(x_2, v, G) \label{eq:tmdd2}
\\&\leq&
2k'+\dist(u,x_2, G)+2(t_v+j)+(2t-1) \cdot  \dist(x_2, v, G)\label{eq:tmdd3}
\\&\leq&
2(k-1)+\dist(u,x_2, G)+(2t-1) \cdot \dist(x_2, v, G) \label{eq:tmdd4}
\\&\leq&
k\cdot \dist(u,x_2, G)+(2t-1) \cdot \dist(x_2, v, G)
\\&\leq&
k \cdot \dist(u, v, G)~, \nonumber
\end{eqnarray}
where Eq. (\ref{eq:tmdd1}) follows by Eq. (\ref{eq:boundpi}),
Eq. (\ref{eq:tmdd2}) follows by Lemma \ref{cl:clut_edge} and by the fact that $u$ is $k'$-clustered and $x_2$ is $t_v+j$ clustered,
Eq. (\ref{eq:tmdd3}) follows by the fact that $w$ and $z$ are the closest pair in $C_{k'}(u)$ and $C_{t_v+j}(x_2)$, finally
the penultimate inequality holds as $\dist(u, x_2, G)\geq 2$ (since $u \neq x_1\neq x_2$).

\paragraph{Case (a2): $u$ is $(k-1-(t_v+j))$-unclustered.}
Let $k'=k-1-(t_v+j)$. Since all vertices are $0$-clustered, it holds that $k'\geq 1$ and hence $j\leq k-t_v-2$. By the definition of $j$, we have that $x_2$ is $(t_v+j+1)$-unclustered (as $j \leq k-t_v-2$, it holds that $(t_v+j+1) \leq k-1$).
Also, recall that $x_1 \neq u$.
Let $\pi(u,v)=[u=u_0, u_1, \ldots, u_q=v]$.
Since $u_0$ is $k'$-unclustered, by Lemma \ref{cl:clut_edge}, it holds $\dist(u_0, u_1,H)\leq 2k'-1$.
In the same manner, since $x_2$ is $(t_v+j+1)$-unclustered, it holds that $\dist(x_1, x_2, H) \leq 2(t_v+j+1)-1$.
Putting all together, we have that
\begin{eqnarray}
\dist(u, v, H) &=&\dist(u_0, u_1,H)+\dist(u_1, x_1,H)+\dist(x_1, x_2, H)+\dist(x_2, u_q,H) \nonumber
\\&\leq&
2k'-1+(2t-1)\cdot \dist(u_1, x_1,G)+2(t_v+j+1)-1 \label{eq:tmpddd1}
\\&+&(2t-1)\cdot \dist(x_2, u_q,G) \nonumber
\\&=&
2(k-1)+(2t-1)\cdot (\dist(u_1, x_1,G)+\dist(x_2, u_q,G)) \nonumber %\label{eq:tmpddd2}
\\&=&
(k-1) \cdot (\dist(u_0, u_1,G)+\dist(x_1, x_2, G))
\\&+&
(2t-1)\cdot (\dist(u_1, x_1,G)+\dist(x_2, u_q,G) \nonumber
\leq
k\cdot \dist(u, v, G) \nonumber~,
\end{eqnarray}
where Eq. (\ref{eq:tmpddd1}) follows by using Lemma \ref{cl:clut_edge} for the $k'$-unclustered edge $(u_0,u_1)$ and  for the $(t_v+j+1)$-unclustered edge $(x_1, x_2)$, and by plugging Eq. (\ref{eq:boundpi}). Note that it might be the case that $x_1=u_1$ but by the current case it holds that $u_0 \neq x_1 \neq x_2$.
\paragraph{Case (b): $\dist(x_1, x_2,G)=1$ and $\dist(u, x_2,G)\geq  2k$.}
Using Part (1) of this claim, we have that
\begin{eqnarray}
\label{eq:tmpall}
\dist(u, v, H) &=&\dist(u, x_1 ,H)+\dist(x_1, x_2, H)+\dist(x_2, v,H)
\\&\leq&
(2t-1) \cdot \dist(u, x_1 ,G)+\dist(x_1, x_2, H)\nonumber
\\&+&
(2t-1)\cdot\dist(x_2, v,G) \label{eq:tmpall1}
\\&\leq&
(2t-1) \cdot \dist(u, x_1 ,G)+2k-1 \nonumber
\\&+& (2t-1)\cdot \dist(x_2, v,G)  \label{eq:tmpall2}
\\&\leq&
(2t-1) \cdot \dist(u, v ,G)+\dist(u, v ,G)\leq 2t \cdot \dist(u, v ,G)~. \label{eq:tmpall3}
\end{eqnarray}
where Eq. (\ref{eq:tmpall1}) follows by Eq. (\ref{eq:boundpi}),
Eq. (\ref{eq:tmpall2}) follows by Part (1) of this claim and Eq. (\ref{eq:tmpall3}) follows by the fact that
$\dist(u, v,G)\geq \dist(u, x_2,G)\geq 2k-1$.
The claim holds.

\paragraph{Case (c): $\dist(x_1, x_2,G)\geq 2$.}
This case follows by using Lemma \ref{cl:tclusvertices}(b) and
Eq. (\ref{eq:boundpi}).
The claim follows.
\QED

\paragraph{Running time.}
Begin with step (S1). By Theorem 3.1 of \cite{BSADD10}, Algorithm $\CLUSTER(G,k,\mu)$ can be implemented in $O(k \cdot m)$ deterministic time. Step (S2) can be implemented in linear time (in the size of the output). Finally, step (S3) requires computing the shortest-path distances between cluster pairs. This can be done by computing the all-pairs-shortest-paths in $O(n^{\omega})$, where $\omega<2.373$ denotes the matrix multiplication exponent.
This completes the proof of Thm. \ref{thm:hybrid}.

%\subsection{Lower Bound}
%\label{subsec:hybrid_lb}
%
%A corollary of the proof of \cite{Woodruff06} is the following. %
%\begin{lemma}
%\label{lem:hybrid_lowerbound}
%Let $1 \leq k \leq O(\ln r/ \ln \ln r)$ for some integer $r \geq 1$. There exists an unweighted undirected graph $G=(V,E)$ with $|V|=\Theta(k r)$ vertices such that for any subgraph $H \subseteq G$ with less than $c \cdot r^{1+\frac{1}{k}})$ edges, for some constant $c \geq 1$, there exists a pair of vertices $u$ and $v$ satisfying that $\dist(u,v,G)=2$ but $\dist(u, v, H)>k \cdot \dist(u, v, G)=2k$.
%\end{lemma}

\section{Sourcewise spanners}
\label{sec:sourcewise}
In this section, we provide several constructions for sourcewise spanners and emulators.
\subsection{Upper bound for multiplicative stretch}
\label{subsec:sourcewise_ub}
For clarity of presentation, we describe a randomized construction whose output spanner has $O(k^2 \cdot n^{1+\varepsilon/k})$ edges in expectation.
Using \cite{BSADD10}, this construction can be derandomized with the same bound on the number of edges.
%In Appendix Sec. \ref{append:hybrid}, we then present a deterministic way of constructing a hybrid spanner with $O(k^2 \cdot n^{1+\varepsilon/k} \cdot \log n)$ edges with the same running time bounds.
%\begin{theorem}
%\label{thm:ub_mult}
%Let $k \geq 1$ be an integer. For every unweighted $n$-vertex graph $G=(V,E)$ with $m$ edges, and every subset $S \subseteq V$, a $(2k-1,S)$ sourcewise spanner $H \subseteq G$ can be constructed in $O(k^2 \cdot m)$ time with $O(k^2 \cdot n^{1+\varepsilon/k})$ edges, where $\varepsilon=\log|S|/\log n$.
%\end{theorem}
We begin by considering Thm. \ref{thm:hybridsw} and show the construction of $(2k-1,S)$ sourcewise spanner which enjoys a ``hybrid" stretch, though in a weaker sense than in Sec. \ref{sec:hybrid}. Specifically, we show that the neighbors of $S$ enjoy a multiplicative stretch $2k-1$ and the remaining pairs enjoy a multiplicative stretch of $2k-2$.
%I.e., $\dist(s_i, v,H) \leq  k' \cdot \dist(s_i, v,G)$ where $k'=2k-1$ if $(s_i, v) \in E(G)$ and $k'=2k-2$ otherwise.
\paragraph{The algorithm}
The first phase of Algorithm $\SOURCEWISES$ applies Algorithm $\CLUSTER(G, k, \mu)$ for $\mu=\varepsilon/k$, resulting in a sequence of $k+1$ clusterings $\mathcal{C}_0, \ldots, \mathcal{C}_{k}$ and a cluster graph $H_{k} \subseteq G$. In the second phase of the algorithm, it considers the collection of $S \times Z_{k-1}$ shortest paths $\mathcal{P}=\{\pi(s_j, z_i) \mid s_j \in S \mbox{~and~} z_i \in Z_{k-1}\}$. Starting with $H=H_{k}$, for each path $\pi(s_j,z_i) \in \mathcal{P}$, it adds to $H$ the $\ell_k$ last edges of $\pi(s_j,z_i)$ (closest to $z_i$). Set
\begin{equation}
\label{eq:nlast}
\ell_k = 2k^2+3k~ \mbox{~~and~~~}\mu=\varepsilon/k~.
\end{equation}
%Appendix Sec. \ref{sec:sw_nh},
%In the full version, we provide a complete analysis for the algorithm and establish Thm. \ref{thm:ub_mult}.
\dnsparagraph{Analysis}
Hereafter, we mainly focus on the clustering $\mathcal{C}_{k-1}$ and denote a vertex $v$ as \emph{clustered} iff $v \in V_{k-1}$, i.e., if $v$ belongs to the clustering $\mathcal{C}_{k-1}$. Recall that each of the $m_{k-1}=n^{1-(k-1) \cdot \mu}$ clusters $C_i \in \mathcal{C}_{k-1}$ is centered at some vertex $z_i \in Z_{k-1}$, and in addition, every vertex $v$ satisfying that $\dist(v, Z_{k-1}, G) \leq k-1$ is contained in exactly one cluster $C_i \in \mathcal{C}_{k-1}$ such that
\begin{equation}
\label{eq:clust_crit}
\dist(v, z_i, G) \leq k-1~.
\end{equation}
An edge $e=(x,y)$ is said to be \emph{clustered} iff both its endpoints $x$ and $y$ are clustered, i.e., $x, y \in V_{k-1}$. By Lemma \ref{lem:sspanner}(1), we have the following.
\begin{observation}
\label{obs:multspanner_size_final}
$\mathbb{E}(|H|)=O(k^2 \cdot n^{1+\varepsilon/k})$.
\end{observation}
%\Proof
%By construction $|H|\leq |H_{k-1}|+\ell_k \cdot |\mathcal{P}|$.
%By Lemma \ref{lem:sspanner}, $\mathbb{E}(|H_{k-1}|)=O(n^{1+\mu})$.
%Since $|\mathcal{P}|=|S| \times |Z_{k-1}|=O(n^{\varepsilon} \cdot n^{1-{k-1} \cdot \mu})$, by Eq. (\ref{eq:nlast}), both $\mathbb{E}(|H_{k-1}|)$ and $|\mathcal{P}|$ are $O(n^{1+\varepsilon/k})$. The observation now follows by Eq. (\ref{eq:nlast}).
%\QED
Next, we turn to consider correctness.
Call a subgraph $H' \subseteq G$ \emph{happy} iff every missing edge $e=(x,y) \in E(G) \setminus E(H')$ is either clustered or has stretch at most $2(k-1)-1$ in $H$, i.e., $\dist(x,y, H) \leq 2k-3$. By Lemma \ref{cl:clut_edge}, we have the following.
%\begin{lemma}
%\label{lem:delta}
%For every $i \in \{1, \ldots, k-1\}$ and for every $v \in \Delta_i$ it holds that $\dist(v,u, H_i) \leq 2i-1$ for every $u \in \Gamma(v,G)$.
%\end{lemma}
%\def\APPENDDELTAUB{
%\Proof
%By induction on $i$.
%For the induction base, $i=1$, consider
%$v \in \Delta_i$. Since every neighbor $u \in \Gamma(v,G)$ is in a distinct cluster in $\mathcal{C}_0$, the entire edge set of $v$ is added to $H_1 \subseteq H_{k-1}$. Next, assume the claim holds up to $i-1\geq 1$ and consider $i$. Let $v \in \Delta_{i}$, and consider some neighbor $u \in \Gamma(v, G)$. There are two cases to consider. (a) $u$ is clustered in $\mathcal{C}_{i-1}$. Then either $(u,v)$ was added to $Q_i$ implying that $\dist(u, v, H_{k-1})=1$, or $(u,v)$ was not added, implying that there exists some other edge $(u',v) \in Q_i$ such that $u'$ and $u$ are in the same cluster in $\mathcal{C}_{i-1}$. Hence, by Obs. \ref{obs:clust_connect}, it holds that $\dist(u, v, H_{k-1}) \leq 1+2(i-1)$. (b) $u$ is not clustered in $\mathcal{C}_{i-1}$. Then $u \in \Delta_{i'}$ for some $i'<i$, hence by the induction assumption $\dist(u,v,H_{k-1})\leq 2i'-1 < 2i-1$. The claim follows.
%\QED
%}%\APPENDDELTAUB
%
\begin{observation}
\label{obs:missing}
$H_{k}$ is happy (i.e., for every edge $e=(x,y) \in G \setminus H_{k}$, either $e$ is clustered or $\dist(x,y, H_{k})\leq 2k-3$).
\end{observation}
We now bound the stretch on $S \times Z_{k-1}$, i.e., between sources and cluster centers.
\begin{lemma}
\label{cl:centers}
For every $s_j \in S$ and $z_i \in Z_{k-1}$:\\
If $\dist(s_j, z_i, G)\leq \ell_k$ then $\dist(s_j, z_i, H)=\dist(s_j, z_i, G)$,
else $\dist(s_j, z_i, H) \leq 2(k-1) \cdot \left(\dist(s_j, z_i, G)+1 \right)-\ell_k$.
\end{lemma}
\Proof
Fix $s_j \in S$ and let the cluster centers $Z_{k}=\{z_1, \ldots, z_{m}\}$ be ordered in \emph{nondecreasing} distance from $s_j$, i.e., $\dist(s_j, z_1, G) \leq \dist(s_j, z_2, G) \leq \ldots \leq \dist(s_j, z_{m}, G)$. (Note that the ordering of the centers plays a role only in the analysis and not in the algorithm itself.)
Assume, towards contradiction, that the claim does not hold, and let $\widehat{z}$ be the first center in the ordering for which the claim does not hold.
We first claim that
\begin{equation}
\label{eq:DG}
D_G=\dist(s_j,\widehat{z},G)= |\pi(s_j, \widehat{z})| > \ell_k~,
\end{equation}
since by construction, if $|\pi(s_j, \widehat{z})| \leq \ell_k$ then $\pi(s_j, \widehat{z})$ is taken entirely into $H$ and $\dist(s_j, \widehat{z},H)=\dist(s_j, \widehat{z},G)$, contradiction.
The contradiction assumption is therefore that $D_H=\dist(s_j, \widehat{z},H)$ satisfies $D_H>2(k-1) \cdot (D_G+1)-\ell_k$.
We now distinguish between two cases depending on the type of missing edges in $\pi(s_j,\widehat{z}) \setminus H$ (by the contradictory assumption such missing edges exist). Case (a) is where none of the missing edges is clustered.
By Obs. \ref{obs:missing},
we have that
$D_H \leq (2k-3) \cdot D_G \leq 2(k-1) \cdot D_G-\ell_k$, where the last inequality follows by Eq. (\ref{eq:DG}).
\par Next, consider the complementary case (b) where some of the missing edges of $\pi(s_j, \widehat{z}) \setminus H$ are clustered. Let $e=(x_1, x_2)$ be the last missing edge on the path $\pi(s_j, \widehat{z})$ (the edge closest to $\widehat{z}$) which is clustered, hence $x_2 \in V_{k-1}$.
Consider the cluster center $z'$ of the cluster $C$ of $x_2$.
We claim that $z'$ precedes $\widehat{z}$ in the ordering.
To see this, recall that the last $\ell_k$ edges of the path $\pi(s_j, \widehat{z})$ were taken into $H$, thus $\dist(s_j, x_2, G) \leq \dist(s_j, \widehat{z}, G)-\ell_k=D_G-\ell_k$.
Combining with Eq. (\ref{eq:clust_crit}),
\begin{equation}
\label{eq:t1}
\dist(s_j,z', G) \leq \dist(s_j, x_2, G)+k-1 \leq D_G-\ell_k+k-1<D_G~,
\end{equation}
where the first inequality follows by Eq. (\ref{eq:clust_crit}) and the strict inequality follows by Eq. (\ref{eq:nlast}). This strict inequality implies that indeed $z'$ precedes $\widehat{z}$ in the ordering and by the definition of $\widehat{z}$ (i.e., the first center that violates the claim) it follows that $z'$ satisfies the lemma.

Consider the following alternative $s_j-\widehat{z}$ path $P=P_1 \circ P_2 \circ P_3$ which consists of three segments:
a source-center path $P_1 \in SP(s_j, z', H)$, an intra-cluster path $P_2 \in SP(z', x_2, H)$ and $P_3 \in SP(x_2, \widehat{z}, H)$.
Since $P \subseteq H$ by definition, it remains to bound its length. By Eq. (\ref{eq:clust_crit}), $|P_2| \leq k-1$. By Obs. \ref{obs:missing}, since $P_3$ is free from missing clustered edges, $|P_3| \leq (2k-1)\cdot \dist(x_2, \widehat{z},G)$. To bound the length of $P_1$ we distinguish between two cases depending on the size of $\pi(s_j, z')$. Case (a): $|\pi(s_j, z')| \leq \ell_k$. Then, $\dist(s_k, z',H)=\dist(s_k, z',G)$ and overall we have that
\begin{eqnarray*}
D_H &\leq & |P|= |P_1|+ |P_2|+|P_3| 
\\&\leq& \dist(s_j, z', G)+k-1+(2k-3)\cdot \dist(x_2, \widehat{z},G)
\\&\leq&
D_G-\ell_k+2(k-1)+(2k-3)\cdot \dist(x_2, \widehat{z},G)
\\&\leq&
2(k-1) \cdot \left(D_G+1 \right)-\ell_k~,
\end{eqnarray*}
where the third inequality follows by Eq. (\ref{eq:t1}). This contradicts the assumption that $\widehat{z}$ violates the claim. Case (b): $\dist(s_j, z',G) \geq \ell_k+1$.
Since $z'$ satisfies the lemma,
\begin{equation}
\label{eq:cc}
\dist(s_j, z', H) \leq 2(k-1) \cdot \left(\dist(s_j, z', G)+1 \right)-\ell_k.
\end{equation}
Overall, we get that
\begin{eqnarray}
\label{eq:mult3}
D_H &\leq& |P|=|P_1|+|P_2|+|P_3|
\\&\leq&
2(k-1) \cdot (\dist(s_j, z', G)+1)-\ell_k \nonumber
\\&+& k-1+(2k-3)\cdot\dist(x_2, \widehat{z},G) \label{eq:m1}
\\&=&
\dist(s_j, z', G) +(2k-3)\cdot\dist(s_j, z', G)+3(k-1)-\ell_k \nonumber
\\&+&
(2k-3)\cdot\dist(x_2, \widehat{z},G) \nonumber
\\& \leq & \nonumber
D_G-\ell_k+k-1+(2k-3)\cdot\dist(s_j, z', G)
\\&+&
3(k-1)-\ell_k+(2k-3)\cdot\dist(x_2, \widehat{z},G) \label{eq:m2}
\\& \leq &
D_G-\ell_k+k-1+
(2k-3)\cdot \left(\dist(s_j, x_2, G)+k-1 \right) \label{eq:m3}
\\&+& \nonumber
3(k-1)-\ell_k+(2k-3)\cdot\dist(x_2, \widehat{z},G)
\\&=&
2(k-1) \cdot D_G-\ell_k~, \nonumber
\end{eqnarray}
where Eq. (\ref{eq:m1}) follows by Eq. (\ref{eq:cc}),
Eq. (\ref{eq:m2}) follows by Eq. (\ref{eq:t1}),
and Eq. (\ref{eq:m3}) follows by the fact that $x_2$ and $z'$ are in the same cluster of diameter $k-1$ (see Eq. (\ref{eq:clust_crit})).
This is again in contradiction to the definition of $\widehat{z}$. The lemma follows.
\QED
We are now ready to bound the $S \times V$ stretch also for vertices $V \setminus Z_{k-1}$. We have the following.

\begin{lemma}
\label{lem:correct_ub}
For every $(s_j, v) \in S \cdot V$, 
$\dist(s_j, v,H) \leq  k' \cdot \dist(s_j, v,G)$ where $k'=2k-1$ if $(s_j, v) \in E(G)$ and $k'=2k-2$ otherwise.
\end{lemma}
\Proof
Fix $s_j \in S$ and assume, towards contradiction, that there exists some vertex $v \in V$ for which the claim does not hold.
We distinguish between two cases. Case (a): all the missing edges on $\pi(s_j, v) \setminus H$ are unclustered. Then, by Obs. \ref{obs:missing}, it follows that $\dist(s_j, v, H) \leq (2k-3)\cdot \dist(s_j, v, G)$. Case (b): there exists a missing edge which is clustered. Let $e=(x_1, x_2)$ be the last such missing edge on $\pi(s_j, v)$ (closest to $v$) and let $z'$ be the cluster center of the cluster $C$ of $x_2$.
Consider an $s_j-v$ path $P=P_1 \circ P_2 \circ P_3$ in $H$ which consists of the following three segments: $P_1 \in SP(s_j, z', H)$, $P_2 \in SP(z', x_2, H)$ and $P_3 \in SP(x_2, v, H)$.
Note that by Eq. (\ref{eq:clust_crit}) and Obs. \ref{obs:missing},
\begin{equation}
\label{eq:p32}
|P_2| \leq k-1 \mbox{~~~and~~~} |P_3| \leq (2k-3) \cdot \dist(x_2, v, G)~.
\end{equation}
Bounding the size of $P_1$ is more involved and requires the following case analysis which depends on $\dist(s_j, z', G)$.
\paragraph{Case a: $\dist(s_j, z', G) \leq \ell_k$.}
In this case, the path $\pi(s_j, z')$ was added in its entirety to $H$ and hence $\dist(s_j,z',H)=\dist(s_j,z', G)$. Combining with Eq. (\ref{eq:p32}) we have that
\begin{eqnarray}
\label{eq:tmp}
\dist(s_j, v,H) \leq |P|&=&|P_1|+|P_2|+|P_3| \nonumber
\\&\leq&
\dist(s_j,z', G)+k-1+(2k-3)\cdot\dist(x_2, v, G)
\\&\leq & \nonumber
\dist(s_j, x_2,G)+(2k-3)\cdot\dist(x_2, v, G)+2(k-1)~.
\end{eqnarray}
We now look at two sub-cases. Sub-case (a1):
$\dist(s_j, x_2,G)\geq 2$, then \\$\dist(s_j, x_2,G)+2(k-1) \leq 2(k-1) \cdot \dist(s_j, x_2,G)$ and hence by Eq. (\ref{eq:tmp}) we conclude that \\$\dist(s_j, v,H)\leq 2(k-1) \cdot\dist(s_j, v, G)$ as required.
The complementary sub-case (a2) is where $\dist(s_j, x_2,G)=1$.
(By definition, $e=(x_1, x_2)$ is missing and hence $s_j \neq x_2$ or $\dist(s_j, x_2,G)>0$).
This sub-case now splits further.
Case (a2i): $x_2 \neq v$, i.e., $\dist(x_2, v, G)\geq 1$, then by Eq. (\ref{eq:tmp}), $|P| \leq 2(k-1)+1+(2(k-1)-1)\cdot\dist(x_2, v, G)\leq 2(k-1)\cdot \dist(s_j, v, G)$.
Case (a2ii): $x_2=v$ (i.e., $\dist(x_2, v, G)=0$). Then $|P|\leq \dist(s_j, x_2,G)+2(k-1)$, in contradiction to the fact the $v$ violates the claim.
\paragraph{Case b: $\dist(s_j, z', G) > \ell_k$.}
By Lem. \ref{cl:centers} we have that
$|P_1|\leq 2(k-1) \cdot (\dist(s_j, z', G)+1)-\ell_k$.
Overall, by plugging Eq. (\ref{eq:p32}), we have that
\begin{eqnarray}
\dist(s_j, v,H)&\leq &|P|=|P_1|+|P_2|+|P_3| \nonumber
\\&\leq&
2(k-1)\cdot (\dist(s_j, z', G)+1)-\ell_k+k-1 \nonumber
\\&+&
(2k-3)\cdot\dist(x_2, v,G) \nonumber
\\&\leq&
2(k-1) \cdot (\dist(s_j, x_2, G)+k-1)+3(k-1)-\ell_k \nonumber \\&+&(2k-3)\cdot\dist(x_2, v,G) \nonumber %\label{eq:tt1}
\\&\leq& 
2(k-1) (\dist(s_j, x_2, G)+\dist(x_2, v,G)) \nonumber
\\&=&
2(k-1) \cdot\dist(s_j, v,G)~ \nonumber%\label{eq:tt2}~,
\end{eqnarray}
where the penultimate inequality follows by Eq. (\ref{eq:clust_crit}) and the last inequality follows by plugging Eq. (\ref{eq:nlast}), in contradiction to the definition of $v$. The lemma follows.
\QED
We next analyze the running time and establish Thm. \ref{thm:ub_mult}.
\paragraph{Running time.}
The operation of the algorithm is composed of $k$ iterations. In each iteration $\tau$, a cluster graph is constructed in $O(m)$ time (as in \cite{BSADD10}).
In  the second phase of the algorithm, the $s_i-z_j$ paths of length at most $\ell_k$ edges are added to the spanner. This can viewed as constructing cluster graph of radius $\ell_k$, but in this case, the clusters may not be disjoint (i.e., a vertex $v$ belongs to the cluster of $z_j$ if $\dist(v, z_j, G) \leq \ell_k$). This can be implemented in $O(\ell_k \cdot m)$ time.
%Having $Z_t$, to construct the clustering $\mathcal{C}_t$ and the corresponding forest $F_t$, the algorithm should first compute the distances $\dist(v, Z_t,G)$. This can be done in $O(m \cdot \log n)$ time, following \cite{TZ05}, in the following manner.  Add to $G$, a new vertex $s^*$ and edges $e_i=(s^*, z_i)$ of weight $i$ (where $i$ is assumed to be the index of the cluster center $z_i$). Set the weight of all the edges in $G$ to $2n$. Let the modified graph be denoted by $G'$. Compute the distance from the new source $s^*$ to all other vertices in the graph. These distance can be computed efficiently in $O(m)$ time using the SSSP algorithm of Thorup \cite{TSSSP99}, if one assumes that integer weights can fit in one word or in $O(m \log n)$ time otherwise. It is easy to compute $\dist(v, Z_t,G)$ based on $\dist(s^*,v, G')$. Furthermore, the shortest paths tree constructed by applying Thorup algorithm, supplies a witness $z^* \in Z_t$ such that $\dist(v, Z_t, G)=\dist(v, z^*.G)$ and since we impose a priority based on the identity of the cluster center, it holds that short-path ties are decided in a consistent manner. The total construction of the $F_t$ forest can be done in time $O(m)$.
\def\APPENDUBDERAND{
\paragraph{Derandomization.}
The randomized algorithm for constructing the $(2k, S)$-hybrid sourcewise spanner can be derandomized, with only a small loss in the size of the spanner (i.e., adding an extra $\log n$ factor). Randomization is only used in the selection of cluster centers $Z_0 \subseteq Z_1 \subseteq \ldots \subseteq Z_{k-1}$.  A deterministic way of constructing these samples is by using the following lemma of \cite{TZ05}, which is a light modification of Theorem 2.2 of Alon and Spencer \cite{AlonBook}.
\begin{lemma}[Lemma 3.6 of \cite{TZ05}]
\label{lem:alon}
Let $N_1, \ldots, N_n \subseteq U$ be a collection of sets with $|U|=u$ and $|N_t| \geq s$ for every $1 \leq t \leq n$. Then, a set $A$ of size at most $u /s \cdot (\log n+1)$ such that $N_t \cap A \neq \emptyset$, for every $1 \leq t \leq n$, can be found, deterministically, in $O(u +\sum |N_i|)$ time.
\end{lemma}
The sets $Z_0, \ldots, Z_{k-1}$ are constructed one by one.
The set $Z_0$ is simply $V$. Suppose $Z_{t-1}$, for some $t \in \{1, \ldots, k\}$, was already constructed and let $\mathcal{C}_{t-1}$ be the corresponding clustering. Recall that $\widehat{V}_{t-1}=\bigcup_{C \in \mathcal{C}_{t-1}} C$ is the set of vertices that appear in some cluster in the clustering $\mathcal{C}_{t-1}$. Let $X_t= \cap_{j=0}^{t-1} \widehat{V}_{j}$ be the set of vertices that appear in each of the $\mathcal{C}_j$ clustering for every $j \in \{0, \ldots, t-1\}$.
For each vertex $v \in X_t$, define $N_t(v) \subseteq Z_{t-1}$ as the cluster centers in $Z_{t-1}$ whose cluster is adjacent to $v$.
Formally, letting $C_\ell \in \mathcal{C}_{t-1}$ be the cluster whose center is $z_\ell \in Z_{t-1}$, then,
$$N_t(v)=\{ z_{\ell} ~\mid~ C_{\ell} \cap \Gamma(v) \neq \emptyset\}~.$$
Let $\Delta_t=\{v \in X_t ~\mid~ |N_t(v)| \leq n^{\mu} \cdot (\log n+1)\}$ be the vertices $v$ of $X_t$ whose neighborhood $N_t(v)$ is sufficiently small. Then, for every vertex $v \in \Delta_t$ and every cluster $C \in \mathcal{C}_{t-1}$, pick one vertex $u \in \left(C \cap \Gamma(v)\right)$, and add the edge $(u, v)$ to $Q_t$ (hence, adding exactly $|N_t(v)|$ edges for every $v \in \Delta_t$).
Next, consider the vertices in $R_t=X_t \setminus \Delta_t$. By definition, $|N_t(v)|> n^{\mu} \cdot (\log n+1)$ for every $v \in R_t$. Use Lemma \ref{lem:alon} to choose a set $Z_t$ that \emph{hits} all the sets $N_t(v)$, for every $v \in R_t$. Such a set is constructed in time $O(m)$, since every vertex appears in at most one $R_t$ for $t \in \{1, \ldots, k\}$, and hence $\sum_{v \in R_t} N_t(v)\leq m$).
Note that by construction, the vertices of $R_t$ are clustered in $\mathcal{C}_t$ (i.e., the clustering induced by $Z_t$), hence $\Delta_t$ has the same meaning as the in the randomized construction. In addition, since $\bigcup N_t(v)=Z_{t-1}$ and every $N_t(v)$ is of size at least $n^{\mu} \cdot (\log n+1)$, by Lemma \ref{lem:alon}, it holds that $|Z_t| \leq |Z_{t-1}|/n^{\mu}$. Since $Z_0=V$, it holds that $|Z_t|=n^{1-t\cdot \mu}$ for every $t \in \{0, \ldots, k\}$.

\par The correctness follows the exact same line as in the randomized algorithm. It remains to bound the size of the spanner. First, consider the number of edges contributed by the vertices of $\Delta_\tau$ for every $\tau \in \{1, \ldots, k-1\}$. By definition, the $\Delta_\tau$ sets are disjoint, and in addition, for every vertex $v \in \Delta_\tau$, we add $|N_\tau(v)| \leq n^{\mu}(\log n+1)$ edges to $Q_\tau$. So, $\sum |Q_\tau|=O(n^{1+\mu} \cdot \log n)$. Finally, $O(\ell_k \cdot |S| \times |Z_{k-1}|)$ edges are added due to $S \times Z_{k-1}$ short $s_i-z_j$ paths. Note that thanks to Lemma \ref{lem:alon}, the deterministic running time is $O(k^2 \cdot m)$ as in the randomized setting. This completes the analysis for the deterministic construction.
}%\APPENDUBDERAND

\subsection{Lower bound for additive sourcewise spanners and emulators}
\label{sec:lb_add}
We now turn to consider the lower bound side where we generalize the lower bound construction for additive spanners by Woodruff \cite{Woodruff06} to the sourcewise setting. In particular, we parameterize our bound for the $S \times V$ spanner in terms of the cardinality of the source set $S$.
The basic idea underlying Woodruff's construction is to form a dense graph $G$ by gluing (carefully) together many small complete bipartite graphs. For an additive stretch $2k-1\geq 1$, the lower bound graph $G$ consists of $k+1$ vertex levels, each with $O(n/k)$ vertices and $\Omega(n^{1+1/k})$ edges connecting the vertices of every two adjacent levels. In particular this is obtained by representing each vertex of level $i$ as a coordinate in $\mathbb{Z}^{k+1}$, namely, $v=(a_1, \ldots, a_k,a_{k+1})$ and $a_j \in [1, O(n^{1/k})]$. Woodruff showed that if one omits in an additive spanner $H \subseteq G$, an $O(1/k)$ fraction of $G$ edges, then there exists an $x-y$ path $P$ in $G$ of length $k$ (i.e., $x$ is on the first level and $y$ is on the last level) whose all edges are omitted in $H$, and any alternative $x-y$ path in $H$ is ``much" longer than $P$. To adapt this construction to the sourcewise setting, some \emph{asymmetry} in the structure of the $k+1$ levels should be introduced. In the following construction, the vertices of the first level correspond to the source set $S$, hence this level consists of $O(n^{\varepsilon})$ vertices, while the remaining levels are of size $O(n/k)$. This is achieved by breaking the symmetry between the first coordinate $a_1$ and the remaining $k-1$ coordinates of each vertex $v=(a_1, \ldots, a_k,a_{k+1})$. Indeed, this careful minor adaptation in the graph definition is sufficient to generalize the bound, the analysis follows (almost) the exact same line as that of \cite{Woodruff06}.
%To obtain a graph with a superlinear number of edges, the parameters of Woodruff's construction should be carefully adapted (e.g., the rule for an edge between two vertices).
We show the following.
\begin{theorem}
\label{thm:lb_add_ex}
Let $1 \leq k \leq O(\ln r/ \ln \ln r)$ for some integer $r \geq 1$. For every $\varepsilon \in [0,1]$, there exists an unweighted undirected graph $G=(V,E)$ with $|V|=\Theta(r^{\varepsilon}+k r)$ vertices and a source set $S \subseteq V$ of size $\Theta(r^{\varepsilon})$ such that any $(2k-1,S)$-additive sourcewise spanner $H \subseteq G$ has $\Omega(r^{1+\frac{\varepsilon}{k}})$ edges.
Similar bounds (up to factor $O(k)$) are achieved for $(2k-1,S)$-additive sourcewise emulators.
\end{theorem}
Note that Thm. \ref{thm:lb_add_ex} implies Thm. \ref{thm:lb}, since $n= \Theta(r^{\varepsilon}+k r)$ and hence $r^{1+\frac{\varepsilon}{k}}=\Omega(n^{1+\frac{\varepsilon}{k}}/k)$. Note that by setting $\varepsilon=1$, we get the exact same bounds as in Woodruff's construction.
%But now, the size bound of the $(2k-1, S)$-additive sourcewise spanner depends smoothly on the cardinality of the source set $S$.
%\Proof
%The sourcewise construction is a modification of Woodruff's construction \cite{Woodruff06} for the all-pairs additive spanner case.
%(where the additive stretch is bounded for every pair in $V \times V$).
\dnsparagraph{The construction}
Let $N_1=\lceil r^{\varepsilon/k} \rceil$ and $N_2=\lceil (r/N_1^{k-1})\rceil$.
The graph $G$ consists of vertices composed of $k+1$ vertex-levels and connected through a series of $k$ bipartite graphs. Each vertex $v=(a_1, a_2, \ldots, a_{k}, a_{k+1})$ represents a coordinate in $\mathbb{Z}^{k+1}$ where $a_{k+1} \in \{1, \ldots, k+1\}$ is the \emph{level} of $v$.
The range of the other coordinates is as follows.
For every $1 \leq j \leq k$, $a_j \in R_j$, where
$R_1=\{1, \ldots, N_1\}$ if $a_{k+1}=1$ and $R_1=\{1, \ldots, N_2\}$ otherwise.
%\begin{equation*}
%R_1=
%\begin{cases}
%\{1, \ldots, N_1\}, & \text{if $a_{k+1}=1$;}\\
%\{1, \ldots, N_2\}, & \text{otherwise.}
%%\end{cases}
%\end{equation*}
For $j\geq 2$, $R_j=\{1, \ldots, N_1\}$.\\
Edges in $G$ join every level-$i$ vertex
$(a_1, \ldots, a_{i-1},a_i, a_{i+1}, \ldots, a_{k},i)$
to each of the level-$(i+1)$ vertices of the form
$(a_1, \ldots, a_{i-1},c, a_{i+1}, \ldots, a_{k},i+1)$
for every $c \in \{1, \ldots, N_2\}$ if $i=1$ and $c \in \{1, \ldots, N_1\}$ for $i \geq 2$.
Let $L_i=\{(a_1, \ldots, a_{k},i) ~\mid~ a_j \in R_j \mbox{~for~} 1 \leq j \leq k\}$ be the set of vertices on the $i$th level and let $n_i=|L_i|$ denote their cardinality.
Then since $k=O(\ln r /\ln \ln r)$ it holds that
$n_1= N_1^{k} \leq (r^{\frac{\varepsilon}{k}}+1)^{k}
\leq e^{(k+1)/(r^{\varepsilon/k})}=\Theta(r^{\varepsilon})$.
and for every $i \in \{2, \ldots, k+1\}$,
\begin{eqnarray*}
n_i &=& N_2 \cdot N_1^{k-1} \leq (r/N_1^{k}+1)(r^{\frac{\varepsilon}{k}}+1)^{k-1}
\leq
2r^{1-\varepsilon/k}\cdot e^{k/(r^{\varepsilon/k})}
\\&=&
r^{1-\varepsilon/k}\cdot \Theta(r^{\varepsilon/k})=
\Theta(r)~,
\end{eqnarray*}
Overall, the total number of vertices is
$|V(G)|=n_1+k \cdot n_2=\Theta(r^{\varepsilon}+k\cdot r)$.
\par Let $g_i$ be the number of edges connecting the vertices of $L_i$ to the vertices of $L_{i+1}$.
Then $g_1=N_2 \cdot n_1$ and $g_i=N_1 \cdot n_i$ for every $i \in \{2, \ldots, k\}$, thus $g_1=g_2= \ldots=g_{k}$. Hence
$|E(G)|=\sum_{i=1}^{k+1}g_i=k \cdot N_1^{k}\cdot N_2=\Theta(k \cdot r^{1+\varepsilon/k})$.
Let the source set $S$ be the vertex set of the first level, i.e., $S=L_1$, hence $|S|=n_1=\Theta(r^{\varepsilon})$.
%In Appendix Sec. \ref{append:add_lb},
%In the full version, we analyze this graph construction and establish Thm. \ref{thm:lb}.
The analysis of the construction follows the same line as that of \cite{Woodruff06}. For completeness, we briefly describe it.
We begin by  exploring the following \emph{distortion property} on some spanner $ \subseteq G$. Consider any subgraph $H$ with fewer than $E(G)/k=N_1^{k}\cdot N_2=\Theta(r^{1+\varepsilon/k})$ edges.
We now show that there exist vertices $u \in L_1$ and $v \in L_{k+1}$ such that $\dist(u,v,H)\geq \dist(u, v, G)+2k$.
By Lemma 5 of \cite{Woodruff06}, we have the following.
\begin{lemma}
\label{lem:lb_exist}
There exist $k+1$ vertices $v_1, \ldots, v_{k+1}$ such that for each $i$, $v_i \in L_i$ and the edge $(v_{i}, v_{i+1})$ is missing in $H$.
\end{lemma}
%For completeness, the missing proofs of this section, are provided in Appendix Sec. \ref{append:add_lb}.
%\def\APPENDLBEXIST{
%\Proof
%For $i \in \{1, \ldots, k\}$, let $r_i$ be the number of edges in $H$ connecting vertices of $L_i$ to vertices in $L_{i+1}$. Then $\sum_{i=1}^{k+1}r_i <|E(G)|/k$. Choose a vertex $v_1$ in level $1$ uniformly at random. For $2 \leq i \leq k+1$, inductively choose $v_i$ to be a random neighbor of $v_{i-1}$ in $L_{i}$. Each of the edges $(v_i, v_{i+1})$ is then uniformly random. By the union bound
%$$\mathbb{P}[(v_1, v_2), \ldots, (v_{k+1}, v_{k+1}) \mbox{~are missing~}] \geq 1-\sum_{i=1}^{k}r_i/g_i >0.$$
%This holds since $g_i=|E(G)|/k$ for every $i \in \{1, \ldots, k+1\}$ and by the definition of $H$, $|E(H)|=\sum_i r_i < |E(G)|/k$.
%Thus, there exist $v_1, \ldots, v_{k}$ satisfying the conditions of the lemma.
%\QED
%}%\APPENDLBEXIST
Choose $v_1, \ldots, v_{k}$ as in Lemma \ref{lem:lb_exist}. Then $\dist(v_1, v_{k},G) \leq k+1$ since $v_1, \ldots, v_{k+1}$ is a path in $G$. The following lemma shows that $\dist(v_1, v_{k+1},H)$ is large.
%The proof follows immediately by \cite{Woodruff06}, for completeness we provide it here.
\begin{lemma}[Lemma 6 of \cite{Woodruff06}]
\label{lem:lb_large}
Any path in $G$ from $v_1$ to $v_{k+1}$ of length less than $3k$ contains an edge $(v_i, v_{i+1})$ for some $i \in \{1, \ldots, k\}$, and further this is the only path edge on the $i$th level.
\end{lemma}
\Proof
Let $P$ be any path from $v_1$ to $v_{k+1}$ in $G$.
Let $i$ be any level $i \in \{1, \ldots, k+1\}$.
Starting from $v_1$, traverse the edges of $P$ one by one.
After encountering an even number of edges in level $i$ as we walk along $P$, we must be in level $j$ such that $j \leq i$. Thus as $P$ starts with a level-$1$vertex and ends with a level- $(k+1)$ vertex, $P$ must contain an odd number of edges in each level $i$.
\par Therefore, if the length of $P$ is less than $3k$, then by the pigeonhole principle there is an $i$ for which $P$ contains exactly one edge in level $i$ (since otherwise $P$ contains at least 3 edges in each level so $|P| \geq 3(k+1)$).
Let $(a, b)$ denote this edge.
We claim that $(a, b)=(v_i, v_{i+1})$. To see this, first note that the last $k-(i-1)$ coordinates (not including the level coordinate) of $a$ must agree with those of $v_1$ since
(i) $P$ begins at $v_1$, (ii) all edges in $P$ preceding $(a, b)$ are in level $j <i$, and (iii) an edge in level $j$, for any $1 \leq j \leq k$, may only modify the $j$th coordinate of its endpoints. Moreover, as $(a, b)$ is the only edge in level $i$, $P$ cannot return to any level $j <i$. Therefore, since $P$ ends at $v_{k+1}$, the first $i-1$ coordinates of $a$ must agree with those of $v_{k+1}$.
By definition then, $a=v_i$.
As only edges on the $i$th level affect the $i$th coordinate, necessarily $b=v_{i+1}$, as otherwise another edge in level $i$ would be needed to correct the $i$th coordinate so that $P$ could reach $v_{k}$. This proves the lemma.
\QED

\begin{corollary}
\label{lem:lb_3k}
$\dist(v_1, v_{k+1},H) \geq 3k$ hence $\dist(v_1, v_{k+1},H) >\dist(v_1, v_{k+1},G)+2k-1$.
\end{corollary}
\Proof
By the previous lemma, any path in $G$ of length less than $3k$ contains an edge of the form $(v_{i}, v_{i+1})$, and by our choice of $v_1, \ldots, v_{k}$, this edge is missing in $H$ so the path does not occur in $H$.
\QED
It follows that any subgraph of $G$ with less than $|E(G)|/(k+1)=\Omega(r^{1+\varepsilon/k})$ edges distorts some $S \times V$ pair by at least an additive $2k$, so it is not a $(2k-1,S)$-additive sourcewise spanner.
\begin{theorem}
\label{thm:lb_emulator}
Let $1 \leq k \leq O(\ln r/ \ln \ln r)$ be an integer.
For every $\varepsilon \in [0,1]$, there exists an unweighted undirected graph $G=(V,E)$ with $|V|=\Theta(r^{\varepsilon}+k r)$ vertices and a source set $S \subseteq V$ of size $\Theta(r^{\varepsilon})$ such that any $(2(k-1),S)$-additive emulator $H=(V,F)$ has $\Omega(1/k \cdot r^{1+\varepsilon/k})$ edges.
\end{theorem}
The graph $G$ that achieves the lower bound is the same as in the proof of Thm. \ref{thm:lb_add_ex}. The correctness of the construction follows immediately from the proof of Thm. $8$ in \cite{Woodruff06}. In short, one can show that the diameter of $G$ is at most $O(k)$. In addition, note that for any edge $(u,v) \in E(H)$ it is never optimal for the weight of $(u,v)$ to be larger than $\dist(u,v, G)$. Combining these two observation implies that any weighted edge $(u,v)$ in the emulator $H$  can be replaced by at most $O(k)$ edges in $G$, corresponding to the $u-v$ shortest-path in $G$. This results in a \emph{subgraph} $H'$ of $G$ which according to Thm. \ref{thm:lb_add_ex} is of size  $\Omega(1/k \cdot r^{1+\varepsilon/k})$. It follows that the emulator $H$ is sparser by a factor of at most $O(k)$, i.e., $|E(H)|=\Omega(1/k^2 \cdot  r^{1+\varepsilon/k})$, as required.
%%%%%%%%%%%%%%%%%%%%%%
\subsection{Upper bound for additive sourcewise spanners and emulators}
%%%%%%%%%%%%%%%%%%%%%%
\label{sec:up_add}
\paragraph{Additive sourcewise emulators.}
Recall that an emulator $H=(V,F)$ for graph $G$ is a (possibly) weighted graph induced on the vertices of $G$, whose edges are not necessarily contained in $G$.
In Thm. \ref{thm:lb_add_ex}, we showed that every $(2,S)$-additive sourcewise emulator for a subset $S \subseteq V$ has $\Omega(n^{1+\varepsilon/2})$ edges, where $|S| =O(n^{\varepsilon})$. We now show that this is essentially tight (up to constants).
\begin{theorem}
\label{thm:emult}
For every unweighted $n$-vertex graph $G=(V,E)$ and every subset $S \subseteq V$, there exists a (polynomially  constructible) $(2,S)$-additive sourcewise emulator $H$ of size $O(n^{1+\varepsilon/2})$ where $\varepsilon=\log |S| /\log n$.
\end{theorem}
The following Fact from \cite{CGK13} is useful in the subsequent constructions.
\begin{fact}[\cite{CGK13}]
\label{fc:clustering}
There is a polynomial time algorithm $\CLUSTER(G, \gamma)$ that given a parameter $\gamma \in [0,1]$ and a graph $G=(V,E)$ constructs a collection of clusters $\mathcal{C}$ with at most $n^{1-\gamma}$ vertex-disjoint clusters, each of size $n^{\gamma}$, and a subgraph $G_{C}$ of $G$ with $O(n^{1+\gamma})$ edges such that (1) for any missing edge $(u,v) \in E(G) \setminus E(G_C)$, $u$ and $v$ belong to two different clusters and (2) the diameter of each cluster (i.e., the distance in $G_C$ between any two vertices of the cluster) is at most $2$.
\end{fact}
\dnsparagraph{Proof of Theorem \ref{thm:emult}}
The weighted graph emulator $H=(V,F)$ is constructed as follows. Let $W: F \to \mathbb{R}$ be the weights of $H$ edges. Apply Fact \ref{fc:clustering} to construct a clustering $\mathcal{C}$ and the clustering graph $G_C=\CLUSTER(G, \gamma)$ for $\gamma=\varepsilon/2$.
For every source $s_i \in S$ and cluster $C_j \in \mathcal{C}$, define $z_{i,j}  \in C_j$ to be the closest vertex in cluster $C_j$ to the source $s_i$, i.e., $z_{i,j} \in C_j$ satisfies that $\dist(s_i, z_{i,j},G)=\dist(s_i, C_j, G)$, and add to $H$ an edge $e_{i,j}$ between $s_i$ and $z_{i,j}$ of weight $\dist(s_i, z_{i,j},G)$.
This concludes the construction of $H$.
We first bound the size of $H$.
By Fact \ref{fc:clustering}, $G_C$ has $O(n^{1+\varepsilon/2})$ edges and there are $|\mathcal{C}|=O(n^{1-\varepsilon/2})$ clusters. Hence, overall $H$ consists of $|E(G_C)|=O(n^{1+\varepsilon/2})$ edges of weight $1$ and $|S| \cdot |\mathcal{C}|=O(n^{1+\varepsilon/2})$ weighted edges as required.
We now analyze the stretch. Consider some $s_i-v_j$ shortest path $\pi(s_i, v_j)$. If $\pi(s_i, v_j) \subseteq G_C$, then $\dist(s_i, v_j, H)=\dist(s_i, v_j, G)$ and the claim holds. Else, let
$e=(w_1, w_2)$ be last missing edge on $\pi(s_i, v_j) \setminus G_C$, i.e., the edge closest to $v_j$ that does not appear in the clustering graph $G_C$. By Fact \ref{fc:clustering}, $w_2$ is clustered.
Let $C=C(w_2)$ be the cluster of $w_2$ and let $z \in C$ be the closest vertex to $s_i$ in the cluster $C$.
By construction, $H$ contains an edge $e'=(s_i, z)$ of weight $W(e')=\dist(s_i, z, G)=\dist(s_i, C, G)$. Let $P_1=e'$, $P_2 \in SP(z, w_2, G_C)$, $P_3=\pi(s_i, v_j)[w_2, v_j]$ and define the $s_i-v_j$ path $P=P_1 \circ P_2 \circ P_3$. By construction $P \subseteq H$. Let $W(P)=\sum_{e \in P} W(e)$ be weight length of $P$.
Note that by Fc. \ref{fc:clustering}, $W(P_2)\leq 2$.
We then have the following.
\begin{eqnarray*}
\dist(s_i, v_j, H) &\leq& W(P)=W(P_1)+W(P_2)+W(P_3)
\leq
\dist(s_i, C, G)+2
\\&+&\dist(w_2, v_j, G) \leq
\dist(s_i, w_2, G)+2 +\dist(w_2, v_j, G)
\\&=&|\pi(s_i, v_j)|+2=\dist(s_i, v_j, G)+2~.
\end{eqnarray*}
where the last inequality follows by the fact that $w \in C$. The theorem follows.
\QED

\paragraph{Additive sourcewise spanners.}
The construction of additive sourcewise spanners combines the path-buying technique of \cite{BSADD10,CGK13,PairwiseICALP13} and the $4$-additive spanner techniques of \cite{4Add}.
\begin{theorem}
\label{thm:add_sourcewise}
Let $k\geq 1$ be an integer. For every unweighted $n$-vertex graph $G=(V,E)$ and every subset $S \subseteq V$, there exists a (polynomially  constructible) $(2k,S)$-additive sourcewise spanner $H \subseteq G$ of size $\widetilde{O}(k \cdot n^{1+(k\varepsilon+1)/(2k+2)})$ where $\varepsilon=\log |S|/\log n$.
\end{theorem}
We first provide some high level overview of the algorithm.
Inspired by \cite{4Add}, the algorithm handles separately ```distant" vertex pairs and ``nearby" vertex pairs in $S \times V$, where the classification is based on some distance function $\dist^*(s, v, G)$. To provide a bounded stretch for the ``distant" vertex pairs (i.e., whose shortest path in $G$ is long according to the function $\dist^*$), the algorithm picks a small sample of vertices for rooting BFS trees. It is shown that this sample covers with high probability the neighborhood of the paths $\pi(s, v)$, when $s$ and $v$ are distant.
Next, the path-buying procedure of \cite{CGK13} is applied on the collection of the ``short" $s-v$ paths (i.e., handling the nearby vertex pairs in $S \times V$). In the analysis, we show that since the path-buying procedure cares only for the ``short" $\pi(s, v)$ paths, the obtained spanner is sparser than that of \cite{CGK13}.
%In Appendix Sec. \ref{append:add_source}, we provide a full description of the algorithm and establish Thm. \ref{thm:add_sourcewise}.

\paragraph{The algorithm.}
We begin by briefly outline the strategy of the path buying procedure. In an initial clustering phase, a suitable clustering of the vertices is computed and an associated subset of edges is added to the spanner. This is followed by a path-buying phase, which examines certain paths in sequence, and decides whether or not to add them to the spanner. The decision is determined by assigning each candidate path a \emph{cost}, corresponding to the number of path edges not already contained in the spanner, and a \emph{value}, measuring how much adding the path would help to satisfy the considered set of constraints on the pairwise distances. The candidate path is added if the {\em value to cost ratio} is sufficiently large. The path-buying strategy was employed in the context of pairwise spanners both in \cite{CGK13} and \cite{PairwiseICALP13}.
\par The following notation is useful in our setting.
A pair $(s_i, v_j) \in S \times V$ is \emph{satisfied} by a subgraph $H \subseteq G$ if $\dist(s_i, v_j, H) \leq \dist(s_i, v_j, G)+2k$.
The neighborhood of a path $P$ is denoted by $\Gamma(P)=\bigcup_{u \in P}\Gamma(u)$.
\par Define
\begin{equation}
\label{eq:Y}
Y=n^{(k\varepsilon+1)/(2k+2)} \mbox{~~and~~} L=n \log n /Y^2~.
\end{equation}
A vertex $v$ is \emph{heavy} iff $\deg(v,G)\geq Y$, otherwise it is \emph{light}. We now classify the $s_i-v_j$ paths $\pi(s_i, v_j)$ according to the number of heavy vertices that appear on the path.
For a path $P$, let $\dist_{heavy}(P)$ be the number of heavy vertices in $P$.
The path $P$  is \emph{long} if
$\dist_{heavy}(P) \geq L$, otherwise it is \emph{short}. The pair $\langle s_i ,v_j \rangle \in S \times V$ is \emph{long} (resp., \emph{short}) iff $\pi(s_i, v_j)$ is long (resp., short).
Define the subgraph
\begin{equation}
\label{eq:long}
E(H_0)=\{(u,v) \in E(G) ~\mid~ \mbox{~u is light~}\}~.
\end{equation}
as the set of all edges adjacent to light vertices in $G$.
The algorithm consists of two phases that add edges to $H_0$. The goal of the first phase is to satisfy the long pairs and the goal of the second phase is to satisfy the short pairs in the final spanner.
\dnsparagraph{(1) Satisfying Long Pairs}
Initially set $H^a=H_0$. Randomly select a set of vertices $Z$ of expected size $9Y$, by choosing every vertex from $V$ independently at random with probability $9Y/n$.  For every vertex $z \in Z$, construct a BFS tree $BFS(z, G)$ rooted at $z$ spanning all vertices $V$, and add the edges of $BFS(z,G)$ to $H^a$.
\dnsparagraph{(2) Satisfying Short Pairs}
This phase consists of two steps: a clustering step and a path-buying procedure.
\dnsparagraph{(2.1) Clustering}
Apply the clustering procedure of Fact \ref{fc:clustering} with $\gamma=\log_n Y$ which constructs a clustering $\mathcal{C}=\{C_1, \ldots, C_{\lambda}\}$ and a clustering subgraph $G_C$ such that $|E(G_C)|=O(n \cdot Y)$ and $\lambda=O(n/Y)$.
\dnsparagraph{(2.2) Path-buying}
The following path-buying procedure is very similar to the construction of additive sourcewise spanner in \cite{CGK13}. The only modification is that in the current setting, the algorithm iterates over the \emph{short} pairs $\langle s_i ,v_j \rangle$ (as the long pairs are handled in the first phase) and not over all $S \times V$ pairs as in \cite{CGK13}. This fact enables the slight improvement in the size of the spanner.
\par Define $\mathcal{P}_{short}=\{\pi(s_i, v_j) ~\mid~ (s_i, v_j) \mbox{~is short~}\}$ as the collection of short $s_i-v_j$ paths which are the candidates to be bought and added to the spanner and initialize $H^b_0\gets G_C \cup H_0$.
Iterate over all paths $\pi(s_i, v_j) \in \mathcal{P}_{short}$. At iteration $t \geq 0$, given the current spanner $H^b_{t-1}$, consider the $t$'th path in  $\mathcal{P}_{short}$, let it be $\pi(s_i, v_j)$. Define paths $\pi_{i,j}^\ell$ for $0 \leq \ell \leq k$, maintaining the following invariants:\\
(i) $\pi_{i,j}^\ell$ is a path between $s_i$ and $v_j$ of length at most $\dist(s_i, v_j, G)+2\ell$. \\
(ii) any cluster $C \in \mathcal{C}$ contains at most three vertices of $\pi_{i,j}^\ell$, \\
(iii) $\Cost(\pi_{i,j}^\ell) \leq L/ \varphi^\ell$ where $\Cost(\pi_{i,j}^\ell)$ is the number of edges of $\pi_{i,j}^\ell$ absent in the current spanner $H^b_{t-1}$ and $\varphi=(2L)^{1/k}$.
\par In the analysis, we show that for every short pair $(s_i, v_j)$, the algorithm buys exactly one path $\pi_{i,j}^\ell$ for $0 \leq \ell \leq k$, hence ensuring that the pair $(s_i, v_j)$ is satisfied in  $H^b_{t}$. We now describe the construction of the $\pi_{i,j}^\ell$ paths.  For $\ell=0$, define
$\pi_{i,j}^0=\pi(s_i,v_j)$, i.e., the shortest-path in $G$. Observe that for $j=0$, Invariant (i) is trivially satisfied, Invariant (ii) is satisfied by Fact \ref{fc:clustering} (otherwise $\pi_{i,j}^0$ would not be a shortest-path), and Invariant (iii) is satisfied because the number of missing edges in $H_0^b$ is at most $L$ (i.e., the pair $(s_i, v_j)$ is short). For a given path $\pi_{i,j}^\ell$, where $\ell \in \{0, \ldots, k\}$, define the function $\Value(\pi_{i,j}^\ell)$ as the number of clusters $C \in \mathcal{C}$ such that there exists a vertex $v \in C \cap \pi_{i,j}^\ell$ and the distance between $s_i$ and $C$ in $H^b_{t-1}$ is strictly greater then the distance between $s_i$ and $C$ in $\pi_{i,j}^\ell$, i.e., $\dist(s_i, C, \pi_{i,j}^\ell)<\dist(s_i, C, H^b_{t-1})$. Formally,
$$\Value(\pi_{i,j}^\ell)=|\{C ~\mid \exists v \in C \cap \pi_{i,j}^\ell ~\mbox{~and~} \dist(s_i, C, \pi_{i,j}^\ell)<\dist(s_i, C, H^b_{t-1})\}|~.$$
The path $\pi_{i,j}^\ell$ is added to $H^b_{t-1}$ resulting in $H^b_{t}$ iff
$$\Cost(\pi_{i,j}^\ell) \leq 3 \cdot \varphi \cdot \Value(\pi_{i,j}^\ell).$$
In other words, if the condition above holds, then $H^b_{t}=H^b_{t-1} \cup \pi_{i,j}^\ell$, the remaining values of $\ell$ are ignored and the algorithm proceeds to the next short pair. Otherwise, the path $\pi_{i,j}^{\ell+1}$ is constructed as follows.
Let $R$ be the longest suffix of $\pi_{i,j}^\ell$ containing exactly
$\lfloor \Cost(\pi_{i,j}^\ell) / \varphi \rfloor$ edges the are missing in $H^b_{t-1}$. By the maximality of $R$, the edge $e$ that precedes $R$ is missing in $H^b_{t-1}$ and hence by Fact \ref{fc:clustering} both endpoints of $e$ (one of which is the first vertex of $R$) are clustered. Hence, at least $1+\lfloor \Cost(\pi_{i,j}^\ell)/\varphi \rfloor \geq \Cost(\pi_{i,j}^\ell)/\varphi$
vertices of $R$ are clustered.
By Invariant (ii) there are at least $\Cost(\pi_{i,j}^{\ell})/(3\varphi)$
clusters in $\mathcal{C}$ having at least one vertex of $R$. Since $\pi_{i,j}^\ell$ was not bought, there exists a cluster $C \in \mathcal{C}$ containing a vertex $x \in C$ of $R$ such that the distance between $s_i$ and $C$ in $H^b_{t-1}$ is at most the distance between $s_i$ and $x$ in $\pi_{i,j}^\ell$.
The path $\pi_{i,j}^{\ell+1}$ is constructed by taking a shortest-path in $H^b_{t-1}$ from $s_i$ to the closest vertex $y \in C$ , then add a path of length at most two between $y$ and $x$ (which exists since $x$ and $y$ belongs to the same cluster and by Fact \ref{fc:clustering}), and finally add the suffix of $R$ starting at $x$.
We now claim that the path $\pi_{i,j}^\ell$ maintains the invariants.
Invariant (i) follows immediately as the length of $\pi_{i,j}^\ell$ was extended by at most two edges (by the selection of $y$ and the addition of the intercluster path of length two). In addition note that $\pi_{i,j}^{\ell+1}$ can be easily transformed into a path which is not longer and satisfies Invariant (ii). In particular as long as the current path contains at least four vertices of the same cluster $C' \in \mathcal{C}$, the path can be shorten as follows. Let $a$ and $b$ be the vertices of the current path closest to $s_i$ and $v_j$ respectively,  replace the $a-b$ subpath (of length at least three) by an the intercluster $a-b$ path in $G_C$ of length at most two. Consequently, Invariant (ii) is satisfied.
Finally, note that the missing edges of $\pi_{i,j}^{\ell+1} \setminus H^b_{t-1}$ (i.e., the edges that contributes to $\Cost(\pi_{i,j}^{\ell+1})$) are fully contained  in $R$. By the definition of $R$, $|R| \leq \Cost(\pi_{i,j}^{\ell})/\varphi^\ell \leq L/(\varphi^\ell \cdot \varphi)$, Invariant (iii) is satisfied. This completes the construction of $\pi_{i,j}^{\ell+1}$.
Let $t'=|\mathcal{P}_{short}|$ be the number of paths considered to be bought. The $(2k, S)$-additive sourcewise spanner $H$ is given by
$$H=H^a \cup H^b_{t'}.$$
\paragraph{Analysis.}
We begin by analyzing the first phase of the algorithm and show that the long pairs are satisfied in $H^a$.
\begin{lemma}
\label{lem:add_long}
(1) With probability at least $1-1/n$, every long pair $(s_i, v_j) \in S \times V$ is satisfied in $H^a$. (2) $H^a$ has $\widetilde{O}(n^{1+(k\varepsilon+1)/(2k+2)})$ edges in expectation.
\end{lemma}
\Proof
Begin with part (1) and consider some long $s_i-v_j$ path $\pi(s_i, v_j)$. Since $\pi(s_i, v_j)$ is a shortest-path in $G$, every vertex $v \in \pi(s_i, v_j)$ has  at most two neighbors in $\pi(s_i, v_j)$.
Combining this with the fact that $\pi(s_i, v_j)$ contains at least $L$ heavy vertices whose sum of degrees is more than $Y \cdot L$, we get that $|\Gamma(\pi(s_i, v_j))|\geq Y \cdot L/3$. We now claim that the probability that $\Gamma(\pi(s_i, v_j)) \cap Z \neq \emptyset$ is at least $1-1/n^3$, as
$$\mathbb{P}[\Gamma(\pi(s_i, v_j)) \cap Z =\emptyset] \leq (1-9Y/n)^{Y \cdot L/3} \leq 1/n^3,$$
which follows by Eq. (\ref{eq:Y}).
Part (1) of the lemma follows by applying the union bound over all pairs of vertices (although it is sufficient to consider only the $S \times V$ pairs).  Consider part (2). The size of $H_0$ is $O(n Y)$ as it consists of edges adjacent to light vertices. The expected number of vertices in $Z$ is $O(Y)$.  For each $z \in Z$, a BFS tree of $n-1$ edges is added to $H$. Hence, the expected number of edges in $H^a$ is $O(n \cdot Y)$. Part (2) follows by plugging Eq. (\ref{eq:Y}).
\QED
%}%\APPENDADDLONG
We now turn to consider the short pairs and analyze the second phase of the algorithm.
\begin{lemma}
\label{lem:add_short}
(1) Every short pair $(s_i, v_j) \in S \times V$ is satisfied in $H^b$. (2) $H^b$ contains $\widetilde{O}(k \cdot n^{1+(k\varepsilon+1)/(2k+2)})$ edges.
\end{lemma}
\Proof
Begin with (1). By Invariant (iii) of the path-buying procedure, we have that $\Cost(\pi_{i,j}^k) \leq 1/2$, since the $\Cost()$ function has only integral values, it has to be that $\Cost(\pi_{i,j}^k)=0$, which ensures that $\pi_{i,j}^\ell$ is bought for some $\ell \leq k$ for every $\pi_{i,j}^0 \in \mathcal{P}_{short}$.
By the above, for every short pair $(s_i, v_j)$, there exists some $\ell \in \{0, \ldots, k\}$ such that $\pi_{i,j}^\ell$ was bought, hence the claim holds by Invariant (i).
Consider part (2).  By Fact \ref{fc:clustering}, $G_C$ contains $n \cdot Y$ edges. Let $\mathcal{B}$ be the paths bought during the path-buying procedure.
It remains to bound the number of edges added due to the paths in $\mathcal{B}$. For every short pair $(s_i, v_j)$ let $P_{i,j}$ be the path in $\mathcal{B}$, i.e., $P_{i,j}=\pi_{i,j}^\ell$ for some $\ell \in \{0, \ldots, k\}$ (by Part (1) above such $\ell$ exists). We first claim that very cluster $C \in \mathcal{C}$ contributes to $\Value(P_{i,j})$ of at most $|S| (2k+3)$ bought paths. This holds since when for $s_i \in S$ a supported path is bought the distance between $s_i$ and $C$ is at most $2k+2$ greater than the distance between $s_i$ and $C$ in $G$: otherwise one could shorten $P_{i,j}$ by more than $2k$, obtaining a contradiction with Invariant (i). Therefore the total number of edges added during the path buying procedure is upper bounded by $\sum_{i,j} \Cost(P_{i,j}) \leq \sum_{i,j} 3 \varphi \cdot \Value(P_{i,j}) \leq 3 \varphi \cdot (2k+3) \cdot |S|\cdot n/Y$. By plugging Eq. (\ref{eq:Y}) and recalling that $|S|=O(n^{\varepsilon})$, the lemma follows.
\QED
Theorem \ref{thm:add_sourcewise} follows by Lemma \ref{lem:add_long} and \ref{lem:add_short}.
\par Finally, we provide an ``almost" tight construction for $(4,S)$-sourcewise additive spanners for a sufficiently large subset of sources $S$. 
We use the following fact due to \cite{CGK13}.
\begin{fact}[Additive subsetwise spanners \cite{CGK13}]
\label{fc:subsetwise}
For every unweighted $n$-vertex graph $G=(V,E)$,
and a subset $Z \subseteq V$, there exists a (polynomially constructible) subgraph $H \subseteq G$ with $O(n^{1+\kappa/2})$ edges such that $\dist(u, v, H) \leq \dist(u, v, G)+2$ for every  $(u, v)\in Z \times Z$, where $\kappa=\log|Z|/\log n$ (this subgraph is also known as \emph{additive subsetwise spanner}).
\end{fact}
%By Theorem \ref{thm:lb_add_ex}, every $(3,S)$-additive sourcewise spanner has $\Omega(n^{1+\varepsilon/2})$ edges where $\varepsilon=\log |S|/\log n$. The next theorem claims that if the set of sources $S \subseteq V$ is sufficiently large, $|S|=\Omega(n^{2/3})$, then there exists a $(4,S)$-additive sourcewise spanner with $O(n^{1+\varepsilon/2})$ edges.
We have the following.
\begin{theorem}
\label{thm:add4sourcewise}
For every unweighted $n$-vertex graph $G=(V,E)$ and a subset of sources $S \subseteq V$ such that $|S|=\Omega(n^{2/3})$, there exists a (polynomially  constructible) $(4,S)$-additive sourcewise spanner $H \subseteq G$ with $O(n^{1+\varepsilon/2})$ edges.
\end{theorem}
\Proof
The algorithm is as follows.
(i) Apply the clustering procedure of Fact \ref{fc:clustering} with $\gamma=\varepsilon/2$. This constructs a clustering $\mathcal{C}=\{C_1, \ldots, C_{\lambda}\}$ and a clustering subgraph $G_C$ such that $|E(G_C)|=O(n^{1+\varepsilon/2})$ and $\lambda=O(n^{1-\varepsilon/2})$.
Let $X=\{x_1, \ldots, x_{\lambda}\}$ be the cluster centers of $\mathcal{C}$. (ii) Set $Z=X \cup S$ and apply the $2$-additive subsetwise spanner construction of Fact \ref{fc:subsetwise} on $Z$, resulting in the subgraph $H'$. Let $H=G_C \cup H'$. Note that $|Z|=O(n^{\varepsilon})$,
since $\varepsilon \geq 2/3$.
Hence by  Facts \ref{fc:clustering} and \ref{fc:subsetwise}, it holds that $H$ has $O(n^{1+\varepsilon/2})$ edges. We now analyze the $S \times V$ stretch in $H$. Consider some $\pi(s_i, v_j)$ path that has some missing edge in $G_C$, i.e., $\pi(s_i, v_j) \setminus G_C \neq \emptyset$. Let $e=(y_1, y_2)$ be the last missing edge on the path (closest to $v_j$). By Fact \ref{fc:clustering}, $y_2$ is clustered. Let $x$ be the cluster center of $C(y_2)$. Hence, $x \in Z$.
Define $P_1 \in SP(s_i, x, H)$, $P_2=(x, y_2)$ and $P_3=\pi(s_i, v_{j})[y_2, v_j]$. Let $P=P_1 \circ P_2 \circ P_3$. It then holds that $P \subseteq H$. By step (ii) it holds that $|P_1| \leq \dist(s_i, x, G)+2$. We have the following.
\begin{eqnarray*}
\dist(s_i, v_j, H) &\leq& |P| =|P_1| +|P_2|+|P_3|
\\&\leq&
\dist(s_i, x, G)+2+1+\dist(y_2, v_j, G)
\\&\leq&
\dist(s_i, y_2, G)+1+3+\dist(y_2, v_j, G)=|\pi(s_i, v_j)|+4
\\&=&\dist(s_i, v_j, G)+4~,
\end{eqnarray*}
where the last inequality follows by the fact the $w$ is the cluster center of $y_2$ and by the triangle inequality.
\QED

\section{Conclusion}
In this paper, we considered the following question:
what is the ``minimal" modification to the standard definition of $(2k-1)$-spanners required in order to bypass the barrier of $\Omega(n^{1+1/k})$ edges, for every integer $k \geq 1$.
We proposed two such modifications. First, we make a hard distinction between pairs at distance $1$ and pairs at distance larger than $1$, corresponding to the hard distinction arising from lower bounds that are based on the girth conjecture. This results in a $k$-hybrid spanner with $O(n^{1+1/(k+1)})$ edges. The $k$-hybrid spanner can be viewed either as a \emph{relaxation} for $k$-spanners (i.e., relaxing the requirement of multiplicative stretch $k$ for neighboring vertex pairs), or alternatively as a \emph{strengthening} of $(\alpha, \beta)$-spanners. Whereas $(\alpha, \beta)$-spanners make a somewhat ``fuzzy" distinction between large and small distances, hybrid spanners provide a \emph{sharp} distinction between vertex pairs at distance $1$ and pairs at distance larger than $1$, which corresponds to the sharp distinction arising by the girth argument. In the second modification, we considered sourcewise spanners, where we care only for the stretch of vertex pairs from $S \times V$  for a given subset of sources $S \subseteq V$. In particular, it follows from our results that even if one considers a large subset of sources $S \subseteq V$ for $|S|=O(n^{1-\delta})$ for sufficiently  small, but fixed $\delta$, there exist a subgraph with $O(n^{1+(1-\delta)/k})$ edges providing a multiplicative stretch of $2k-1$ for every pair $(u, v) \in S \times V$. An interesting future direction is to study the lower bounds for hybrid spanners and sourcewise multiplicative spanners.

\bigskip
\dnsparagraph{Acknowledgment}
I am very grateful to my advisor,
Prof. David Peleg, for many helpful discussions and for
reviewing this paper. I would also like to thank Michael Dinitz and Eylon Yogev for useful comments and discussions.
%\clearpage
{\small

\end{document}